\documentclass[12pt]{article}
\usepackage[colorlinks=true,citecolor=blue,linkcolor=blue]{hyperref}
\usepackage{multirow}% for tables
\usepackage[left=2cm,right=2cm,top=2cm,bottom=2cm]{geometry}
\usepackage{cite}
\usepackage{psfrag}
\usepackage[demo]{graphicx}
\usepackage{graphicx}
\usepackage{graphics}
\usepackage{epsfig}
\usepackage{booktabs}% for tables
\usepackage{amsmath,amsfonts,amssymb}
\usepackage{pstcol,pst-fill,pst-grad}
\usepackage{pstricks,pst-fill,pst-grad}
\usepackage{euscript}
\usepackage{pstricks}
\usepackage{wrapfig}
\usepackage{slashed}
\usepackage[toc,page]{appendix}
\usepackage{here}
\begin{document}
	%\begin{titlepage}
	\title{\vspace{-3cm}
		\hfill\parbox{4cm}{\normalsize \emph{}}\\
		\vspace{1cm}
		{ Laser-assisted charged Higgs pair production in Inert Higgs Doublet Model (IHDM)}}
	\vspace{2cm}
	
	\author{M. Ouali,$^1$ M. Ouhammou,$^1$ S. Taj,$^1$ R. Benbrik,$^2$ and B. Manaut$^{1,}$\thanks{Corresponding author, E-mail: b.manaut@usms.ma} \\
		{\it {\small$^1$ Sultan Moulay Slimane University, Polydisciplinary Faculty,}}\\
		{\it {\small Research Team in Theoretical Physics and Materials (RTTPM), Beni Mellal, 23000, Morocco.}}\\
		{\it {\small$^2$ High Energy Physics and Astrophysics Laboratory, FSSM, UCAM, Morocco.}}\\				
	}
	\maketitle \setcounter{page}{1}
\date{}
\begin{abstract}
The production of a pair of charged Higgs bosons from $e^{+}e^{-}$ annihilation in the presence of a circularly polarized laser field is investigated in Inert Higgs doublet Model (IHDM) at ${e}^{+}{e}^{-}$ colliders. We have derived the analytical expression for the laser-assisted differential cross section by using the Dirac-Volkov formalism at leading order including the $Z$ and $\gamma$ diagrams. Since the laser-free cross section of this process depends, in the lowest order, only on the mass of the charged Higgs boson and the energy of the center of mass, we have analyzed the dependence of the total laser-assisted cross section on these parameters and also on the laser parameters such as the number of photons exchanged, the laser field intensity and its frequency. The results found indicate that the electromagnetic field decreases the order of magnitude of the total cross section as much as the laser field intensity increases or by decreasing its frequency. Moreover, it becomes high for low charged Higgs masses and low center of mass energies.
\end{abstract}
Keywords: Standard model and beyond, Electroweak interaction, Laser-assisted processes, Cross section, IHDM.
\maketitle
   \section{Introduction}
Since its inception in 1960, laser technology has made significant advances, particularly in terms of increasing its intensity and decreasing its pulse duration \cite{1,2}. This advancement have recently received much attention and interest among scientists. This is due not only to the fact that many previously unknown phenomena were induced by the application of an electromagnetic field \cite{3,4,5,6,7} but also for the possibility of testing physical theories such as the standard model and beyond by using powerful laser sources. Several works indicate that laser-matter interaction is relevant in the study of phenomena related to high energy physics such as the standard model of particle physics and beyond, where the electromagnetic field could have a great impact on particles properties and its interactions. In Ref \cite{8}, we have found that the electromagnetic field reduces the total cross section of the Higgs-strahlung production. The same result was obtained in Ref. \cite{9} for neutral Higgs production in Inert Higgs Doublet Model (IHDM). In Ref. \cite{10,11}, authors showed that the circularly polarized electromagnetic field extended the particle's lifetime and enhanced its mode decay.

In the Standard Model (SM) where the implementation of Higgs mechanism with one complet doublet field leads to great discovery of Higgs boson  at the LHC  \cite{12,13}. It also explains the mechanism in which they interact with each other and through which the Higgs-boson generate the mass of all fermions and gauge bosons through Yukawa coupling terms. The physical properties of the new particle \cite{14,15,16,17,18} look compatible with the SM predictions. However, there are many mysterious that SM does not explain, such as, it does not explain the existence of dark matter (DM). Then, many efforts have been made to look for deviations from the SM in the Higgs sector. Therefore, the focus of experimental Higgs searches has shifted toward performing precision measurement in order to establish new evidences for physics beyond the SM \cite{19,20,21,22,23,24}, and to discover other neural or/and charged Higgs bosons, which are present in theories beyond the standard model, where the Higgs sector is extended. 

The  Inert Higgs double model\footnote{which is a special type of general two Higgs doublet (2HDM)} (IHDM) \cite{25,26} is one of the most widely-studied model which is a minimal extension of SM Higgs sector, augmented by a second scalar doublet $H_2$ with vanishing vacuum expectation value (vev). The latter doublet is not concerned by the generation of mass by spontaneous symmetry breaking and do no interact with fermions. In addition, $H_2$ is even under $Z_2$ symmetry. The IHDM Higgs sector contains 5 scalars with at least one neutral Higgs being the SM-like Higgs discovered at the LHC with its mass fixed at 125.1 GeV. The remaining Higgs bosons are one CP-even $(H)$, one CP-odd $(A)$ and two charged Higgs bosons. This model contains also one neutral Higgs which may identified as neutral DM candidate. The combined measurements from both Higgs data and astrophysical observations set a limit on neutral Higgs boson and  wide range of parameters space seems to be excluded \cite{27,28,29}.

Due to free QCD background, an $e^+ e^-$ liner collider is expected to be clean, and as a consequence we do not need any trigger. The physical process is point-like collision at well defined initial energy $\sqrt{s}$ with optimized kinematic conditions where the beams can be polarized. Under such circumstances the final are fully reconstructable and precise cross sections are available. Given that SM has been a great success and its predictions experimentally has been tested with accuracies below than 1\%, it cannot be regarded as the final theory describing the nature. Models accounting for dark matter can be embedded within beyond SM. In this respect, the production of double charged Higgs boson in the presence of a circularly polarized laser field is considered for study at different center of mass energies within the IHDM model at future electron-positron colliders such as Compact Linear Collider (CLIC) \cite{30}, International Linear Collider (ILC),\cite{31,32} FCC-ee and Circular Electron Positron Collider (CEPC)\cite{33,34}. These types of colliders, which possess a very clean environment, will generate precise results and will operate with various collision energies. Thus, the main aim of this research paper is to compute and analyse the total cross section of the charged Higgs boson pair production via electron positron annihilation in IHDM model at various centre of mass energies, and its dependence on the laser field parameters such as the number of exchanged photons, the laser field strength and its frequency. The produced Higgs-bosons are expected to be massive particles, and according to our previous work \cite{7}, the laser field strength required to affect the cross section increases as long as the mass of the particles increases. Therefore, for the intensities considered in this paper, the pair of charged Higgs boson will not interact with the electromagnetic field.

The remainder of the present paper is organized as follows: Section 2 describes the theoretical framework of the charged Higgs-boson pair production process in the presence of a circularly polarized laser field. The obtained numerical results of the total cross section are analyzed and discussed in section 2. A brief conclusion is given in section 3. The expression of  the squared total scattering amplitude is given in the appendix.
\section{Outline of the theory}\label{sec:theory}
In high energy physics, the cross section is one of the most significant measurable quantity that characterizes the scattering process. This part is concerned with the theoretical calculation of the differential cross section for the charged Higgs pair production process via electron positron annihilation ( ${e}^{+}{e}^{-}\rightarrow H^{+}H^{-}$) in IHDM.
The scalar fields of this beyond standard model are complex $SU(2)$ doublet $\Phi_{1}$ and $\Phi_{2}$, with the D-symmetric potential. This implies that only $\Phi_{1}$ can acquire a nonzero vacuum expectation value such that $ v_{1}=<\phi_{1}> $. As a result the scalar fields in $\Phi_{2}$ do not mix with the SM-like field from $\Phi_{1}$.
To obtain a scalar potential that is more closely related to physical observables, one can introduce the so-called Higgs basis in which the redefined doublet fields denoted below by $H^{1}$ and $H^{2}$ have the property that $H^{1}$ has a non-zero vev whereas $H^{2}$ has a zero vev. In particular, we define new Higgs doublet fields such that:
\begin{equation}
H^{1}=
\begin{pmatrix}
G^{\pm}  \\
\frac{1}{\sqrt{2}}(v+h+iG^{0}) 
\end{pmatrix} \hspace*{0.5cm};\hspace*{0.5cm}  \\\ H^{2}= \begin{pmatrix}
H^{\pm}  \\
\frac{1}{\sqrt{2}}(H+iA)
\end{pmatrix},
\label{}
\end{equation}
where, $h$ is SM-like Higgs boson. After electroweak symmetry breaking, the Nambu-Goldstone bosons $G^{\pm}$ and $G^{0}$ are absorbed to give mass to the $W^{\pm}$ and $Z^{0}$ gauge bosons. The dark sector contains four new particles: One neutral scalar $H$, one pseudoscalar $A$ and a pair of charged Higgs $H^{\pm}$. In our previous work \cite{9}, we have studied the neutral Higgs pair production via $e^{+}e^{-}$ annihilation in the presence of a circularly polarized laser field. Therefore, to have a complete picture about the laser field effect on the dark sector, we have investigated the laser-assisted pair production of charged Higgs in IHDM. 
\subsection{Description of the laser field and wave functions}
\begin{figure}[H]
  \centering
      \includegraphics[scale=0.30]{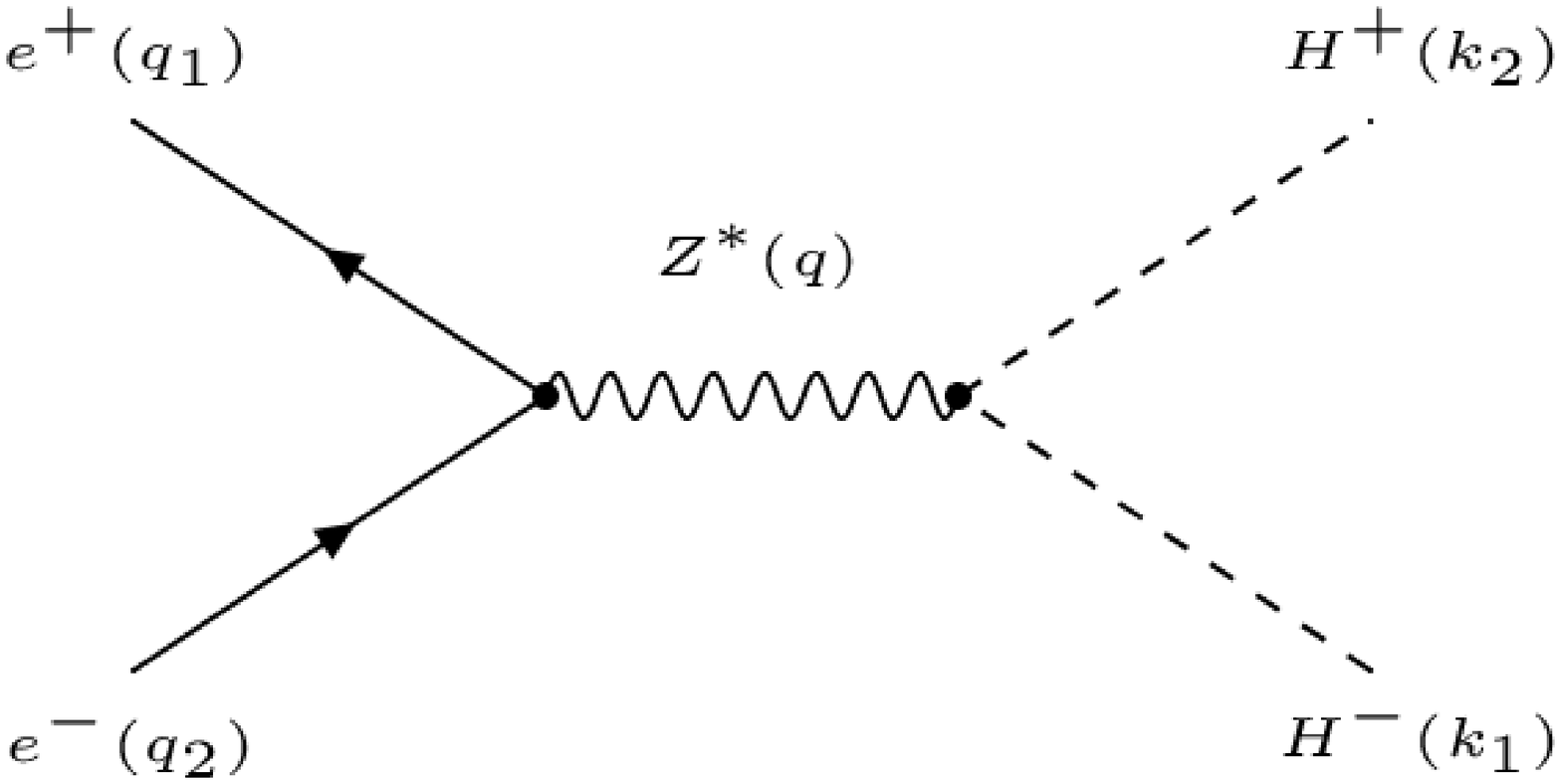}\hspace*{0.4cm}
      \includegraphics[scale=0.30]{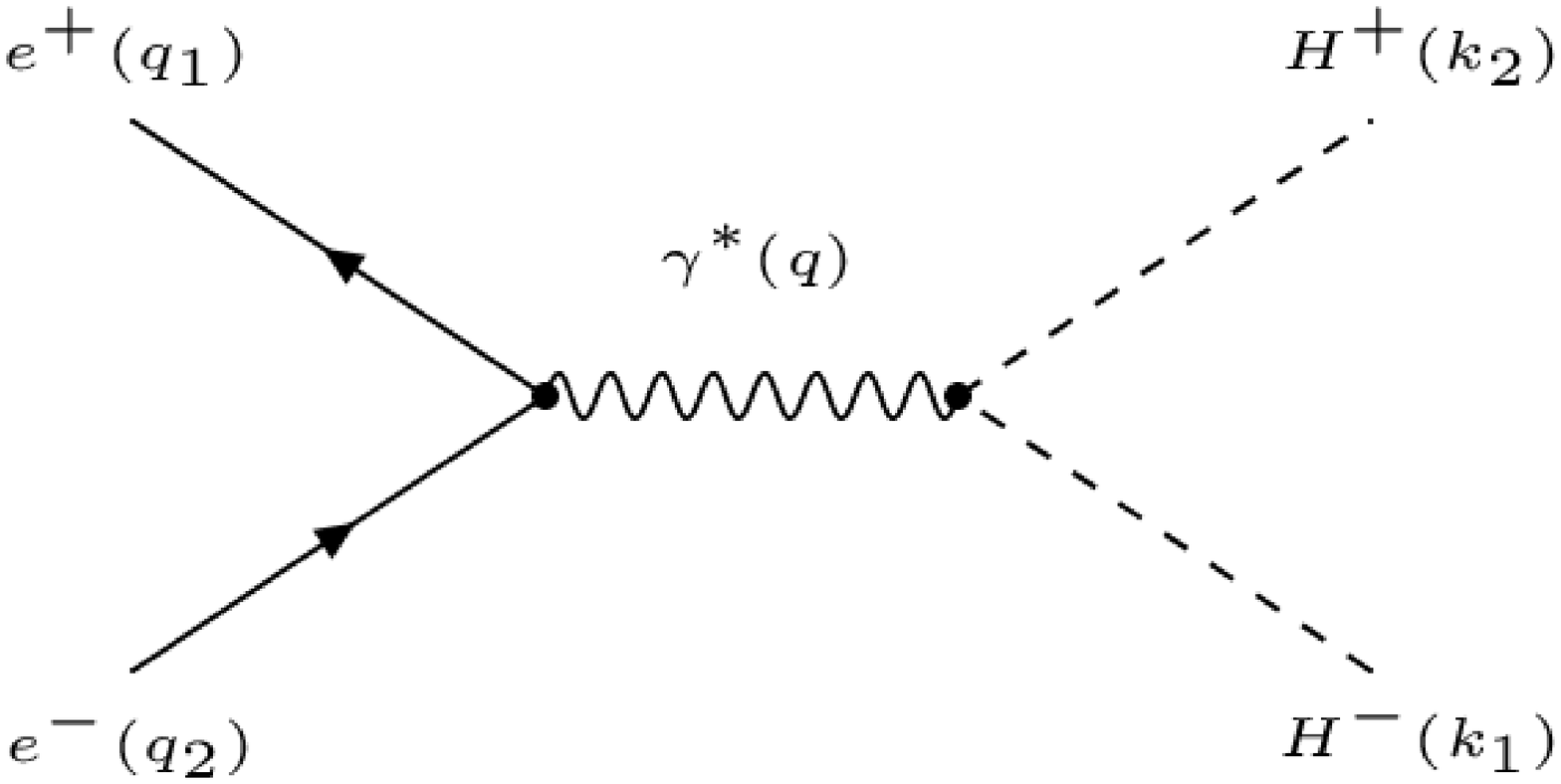}\par\vspace*{0.5cm}
        \caption{Leading-order Feynman diagram  for $s$-channel charged higgs pair production.}
        \label{fig1}
\end{figure}

The process which acts as a source of double Higgs-bosons production at the electron-positron colliders is denoted as: 
\begin{equation}
e^{-}(q_{1}, s_{1})+e^{+}(q_{2}, s_{2})\rightarrow H^{-}(k_{1})+H^{+}(k_{2}),
\label{1}
\end{equation}
where $q_{1}$ is the effective four-momentum of the electron, and $q_{2}$ is the effective momentum of the positron. $k_{1}$ and $k_{2}$ denote successively the free four-momentum of $H^{+}$ and $H^{-}$. $q$ is the four-momentum of the off-shell $\gamma$ and $Z$-boson propagators.
Throughout this paper, we have only considered the leading-order Feynman diagrams which are described by figure \ref{fig1}.
 with the following couplings to charged Higgs boson:
     \begin{equation}
    \gamma H^- H^+ = ie, \quad \quad  ; \quad \quad  Z H^- H^+     = i \frac{g}{2} \frac{\cos(2\theta_W)}{\cos(\theta_W)}.
 \end{equation}
The electron and positron are embedded in a circularly polarized electromagnetic field which is considered to be propagating along the $z$-axis such that its polarization vector is chosen as $k=(\omega,0,0,\omega)$ and $k^{2}=0$, with $\omega$ is the frequency of the laser field. Since we consider intensities which do not allow pair creation, the electromagnetic four-potential is treated classically, and it is given by the following equation:
\begin{equation}
A^{\mu}(\phi)=a_{1}^{\mu}\cos\phi+a_{2}^{\mu}\sin\phi \hspace*{1cm};\hspace*{1cm} \phi=(k.x),
\label{2}
\end{equation}
where $\phi=k.x=k^{\mu}x_{\mu}$ is the phase of the laser field, and $a_{1,2}^{\mu}$ are the polarization four vector, which  are taken as $a_{1}^{\mu}=(0,a,0,0)$ and $a_{2}^{\mu}=(0,0,a,0)$, with $a$ represents the amplitude of the four-potential. $a_{1}^{\mu}$ and $a_{2}^{\mu}$ satisfy the following equations:  $a_{1}^{2}=a_{2}^{2}=a^{2}=-|\mathbf{a}|^{2}=-\big(\epsilon_{0}/\omega \big)^{2}$ where $\epsilon_{0}$ is the strength of the external field. The Lorentz condition $\partial_{\mu}A^{\mu}=0$ implies that $ a_{1}.k=0$ and $\, a_{2}.k=0$. Inside the electromagnetic field, both the electron and positron are described by Dirac-Volkov states \cite{35} such that:
\begin{equation}
\psi_{p_{1},s_{1}}(x)= \Big[1-\dfrac{e \slashed k \slashed A}{2(k.p_{1})}\Big] \frac{u(p_{1},s_{1})}{\sqrt{2Q_{1}V}} \exp^{iS(q_{1},s_{1})},
\label{3}
\end{equation}
\begin{equation}
\psi_{p_{2},s_{2}}(x)= \Big[1+\dfrac{e \slashed k \slashed A}{2(k.p_{2})}\Big] \frac{v(p_{2},s_{2})}{\sqrt{2Q_{2}V}} \exp^{iS(q_{2},s_{2})},
\label{4}
\end{equation}
where, $x$ is the space time coordinate of the incident electron and positron. $u(p_{1},s_{1})$ and $v(p_{2},s_{2})$ are their bispinors,
with $p_{1}(E_{1},|p_{1}|,0,0)$ and $p_{2}(E_{1},-|p_{1}|,0,0)$ refer to their free four-momentum, and 
$ s_{i}(i=1, 2)$ denote their spins. $ Q_{i}(i=1,2)$ is the effective energy acquired by the electron and positron inside the laser field. The arguments of the exponential terms are defined as:
\begin{equation}
S(q_{1},s_{1})=- q_{1}x +\frac{e(a_{1}.p_{1})}{k.p_{1}}\sin\phi - \frac{e(a_{2}.p_{1})}{k.p_{1}}\cos\phi
\label{5}
\end{equation}
\begin{equation}
S(q_{2},s_{2})=+ q_{2}x +\frac{e(a_{1}.p_{2})}{k.p_{2}}\sin\phi - \frac{e(a_{2}.p_{2})}{k.p_{2}}\cos\phi .
\label{6}
\end{equation} 
The effective four-momentum $q_{i}$ is related to its corresponding laser-free momentum $p_{i}$ by the following equation:
\begin{equation}
q_{i}=p_{i}+\dfrac{e^{2}a^{2}}{2(k.p_{i})}k,
\label{7}
\end{equation}
with, $q_{i}^{2}=m_{e}^{*^{2}}=(m_{e}^{2}+e^{2}a^{2})$. $e$ is the electron charge, and $m_{e}$ denotes its mass.
The produced charged Higgs-bosons are massive particles. Thus, for the intensities considered in this paper, the pair of charged Higgs boson are fee, i.e., they will not interact with the laser field. Therefore, they are described by Klein-Gordon states such that:
\begin{equation}
\phi_{k_{1}}(y)=\dfrac{1}{\sqrt{2 E_{H^{-}} V}} e^{-ik_{1}y} \hspace*{0.5cm};\hspace*{0.5cm}  \\\ \phi_{k_{2}}(y)=\dfrac{1}{\sqrt{2 E_{H^{+}} V}} e^{-ik_{2}y},
\label{8}
\end{equation}
where $y$ is the space time coordinate of the outgoing Higgs-bosons. $k_{1}(E_{H^{-}},|k_{1}|\cos \theta,|k_{1}|\sin \theta,0)$ and $k_{2}(E_{H^{+}},-|k_{2}|\cos \theta,-|k_{2}|\sin \theta,0)$ are the free four-momentum of the charged Higgs-bosons, and $E_{H^{-}}$ and $E_{H^{+}}$ denote its corresponding energies.
\subsection{Transition amplitude and cross section}
The leading order scattering-matix element \cite{36} for the laser-assisted charged Higgs-boson pair production in IHDM can be written as:
\small
\begin{eqnarray}
S_{fi}({e}^{+}{e}^{-}\rightarrow H^{+}H^{-})&=&  \int_{}^{} d^4x \int d^4y \Bigg\lbrace\bar{\psi}_{p_{2},s_{2}}(x) \Big( \frac{-ie}{2C_{W}S_{W}  }  \big( \gamma^{\mu} (g_v^{e} -g_a^{e}\gamma^{5}) \big)\Big)    \psi_{p_{1},s_{1}}(x)  D_{\mu\nu}(x-y)  \nonumber \\ &\times &  \phi^{*}_{k_{1}}(y)  \Big( \dfrac{ig \overleftrightarrow{\partial_{\mu}}}{C_{W}}  \big(\frac{1}{2}-S_{W}^{2}\big) \Big)   \phi^{*}_{k_{2}}(y)+\bar{\psi}_{p_{2},s_{2}}(x) (-ie\gamma^{\mu}) \psi_{p_{1},s_{1}}(x)\nonumber \\ &\times & G_{\mu\nu}(x-y)  \phi^{*}_{k_{1}}(y)(ie\overleftrightarrow{\partial_{\mu}})  \phi^{*}_{k_{2}}(y) \Bigg\rbrace,
\label{9}
\end{eqnarray}
\normalsize
where $g_v^{e}$ and $g_a^{e}$ are the vector and axial vector coupling constants, respectively. with $C_{W}=\cos(\theta_{W})$, and $S_{W}=\sin(\theta_{W})$ where $\theta_W$ is the Weinberg angle. {$ g $} is the electroweak coupling constant such that $g^{2}=e^{2}/\sin^{2}\theta_{W}=8G_{F}M_{Z}^{2}\cos^{2}_{\theta_{W}}/\sqrt{2}$. 
The quantity $ D_{\mu \nu}(x-y) $ is the $Z^{*}$-boson propagator, and $G_{\mu \nu}(x-y)$  is the $\gamma^{*}$-boson propagator \cite{36}. They are given by:
\begin{equation}
D_{\mu\nu}(x-y)=\int \dfrac{d^{4}q}{(2\pi)^4} \frac{e^{-iq(x-y)}}{q^{2}-M_{Z}^{2}}\Bigg[-ig_{\mu\nu}+i\dfrac{q_{\mu}q_{\nu}}{M_{Z}^{2}}\Bigg],
\label{10}
\end{equation}
\begin{equation}
G_{\mu\nu}(x-y)=\int \dfrac{d^{4}q}{(2\pi)^4} \frac{e^{-iq(x-y)}}{q^{2}}\Bigg[-ig_{\mu\nu}+i\dfrac{q_{\mu}q_{\nu}}{q^{2}}\Bigg],
\label{11}
\end{equation}
where $q$ denotes their four-momentum.
By substituting the equations (\ref{5}), (\ref{6}), (\ref{8}), (\ref{10}) and (\ref{11}) into the equation (\ref{9}), the scattering matrix element will be as follows: 
\begin{eqnarray}
S_{fi}^{n}({e}^{+}{e}^{-}\rightarrow H^{+}H^{-})&=&\dfrac{(2\pi)^{4}\delta^{4}(k_{1}+k_{2}-q_{1}-q_{2}-nk)}{4V^{2}\sqrt{Q_{1}Q_{2}E_{H^{+}}E_{H^{-}}}} \big(  A_{\gamma}^{n} + A_{Z}^{n} \big).
\label{12}  
\end{eqnarray}
According to the Feynman rules for vector bosons, the total scattering amplitude consists of two parts, which are expressed in terms of Bessel functions as follows:
\begin{eqnarray}
A_{\gamma}^{n}&=& \frac{e^{2}}{(q_{1}+q_{2}+nk)^{2}}  \Bigg\lbrace(k_{2}^{\mu}-k_{1}^{\mu})\bar{v}(p_{2},s_{2})\Bigg[ \chi^{0}_{\mu}\,J_{n}(z)e^{-in\phi _{0}}(z)\nonumber \\ &+ &\frac{1}{2} \,\, \chi^{1}_{\mu}\Big(J_{n+1}(z)e^{-i(n+1)\phi _{0}} + J_{n-1}(z)e^{-i(n-1)\phi _{0}}\Big)\nonumber + \frac{1}{2\, i}\,\chi^{2}_{\mu}\\ &\times& \Big(J_{n+1}(z)e^{-i(n+1)\phi _{0}}-J_{n-1}(z)e^{-i(n-1)\phi _{0}}\Big)\Bigg] u(p_{1},s_{1}) \Bigg\rbrace,
\label{13}
\end{eqnarray}
\begin{eqnarray}
A_{Z}^{n}&=& \frac{e}{2C_{W}S_{W}}  \dfrac{g }{C_{W}}\left(\frac{1}{2}-S_{W}^{2}\right)  \frac{1}{(q_{1}+q_{2}+nk)^{2}-M_{Z}^{2}}  \Bigg\lbrace(k_{2}^{\mu}-k_{1}^{\mu})\bar{v}(p_{2},s_{2}) \nonumber \\ &\times & \Bigg[ \lambda^{0}_{\mu}\,J_{n}(z)e^{-in\phi _{0}}(z)+ \frac{1}{2} \,\, \lambda^{1}_{\mu}\Big(J_{n+1}(z)e^{-i(n+1)\phi _{0}} + J_{n-1}(z)e^{-i(n-1)\phi _{0}}\Big) \nonumber \\ &+& \frac{1}{2\, i}\, \lambda^{2}_{\mu}\Big(J_{n+1}(z)e^{-i(n+1)\phi _{0}}-J_{n-1}(z)e^{-i(n-1)\phi _{0}}\Big)\Bigg] u(p_{1},s_{1})  \Bigg\rbrace,
\label{14}
\end{eqnarray}
where $n$ is the number of photons exchanged between the electromagnetic field and the colliding physical system. The quantities $\chi^{0}_{\mu}$, $\chi^{1}_{\mu}$, $\chi^{2}_{\mu}$, $\lambda^{0}_{\mu}$, $\lambda^{1}_{\mu}$ and $\lambda^{2}_{\mu}$ that appear in equations (\ref{13}) and (\ref{14}) are given by the following expressions:
\begin{equation}
\begin{cases}\chi^{0}_{\mu}=\gamma_{\mu}+2c_{p_{1}}c_{p_{2}}a^{2}k_{\mu}\slashed k
   &\\\chi^{1}_{\mu}=c_{p_{1}}\gamma_{\mu}\slashed k\slashed a_{1}-c_{p_{2}}\slashed a_{1}\slashed k \gamma_{\mu}   
   &\\\chi^{2}_{\mu}=c_{p_{1}}\gamma_{\mu}\slashed k\slashed a_{2}-c_{p_{2}}\slashed a_{2}\slashed k \gamma_{\mu}
    &\\\ \lambda^{0}_{\mu}=\gamma_{\mu}(g_{v}^{e}-g_{a}^{e}\gamma^{5})+2c_{p_{1}}c_{p_{2}}a^{2}k_{\mu}\slashed k(g_{v}^{e}-g_{a}^{e}\gamma^{5})   &\\ \lambda ^{1}_{\mu}=c_{p_{1}}\gamma_{\mu}(g_{v}^{e}-g_{a}^{e}\gamma^{5})\slashed k\slashed a_{1}-c_{p_{2}}\slashed a_{1}\slashed k \gamma_{\mu}(g_{v}^{e}-g_{a}^{e}\gamma^{5})   &\\ \lambda ^{2}_{\mu}=c_{p_{1}}\gamma_{\mu}(g_{v}^{e}-g_{a}^{e}\gamma^{5})\slashed k\slashed a_{2}-c_{p_{2}}\slashed a_{2}\slashed k \gamma_{\mu}(g_{v}^{e}-g_{a}^{e}\gamma^{5}),
\end{cases}
\label{15}
\end{equation}
with $c_{p_{i}}=e/2(kp_{i})$($ i=1,2 $). 
The Bessel function's argument $z$ and its phase $\phi_{0}$ are given by:
$ z=\sqrt{\xi_{1}^{2}+\xi_{2}^{2}}$ and $\phi_{0}= \arctan(\xi_{2}/\xi_{1})$, where:
\begin{center}
$\xi_{1}=\dfrac{e(a_{1}.p_{1})}{(k.p_{1})}-\dfrac{e(a_{1}.p_{2})}{(k.p_{2})}$ \qquad ; \qquad $\xi_{2}=\dfrac{e(a_{2}.p_{1})}{(k.p_{1})}-\dfrac{e(a_{2}.p_{2})}{(k.p_{2})}$.\\
\end{center}
In the centre of mass frame, the differential cross section has the following form:
\begin{equation}
d\sigma_{n}=\dfrac{|S_{fi}^{n}|^{2}}{VT}\frac{1}{|J_{inc}|}\frac{1}{\varrho}V\int_{}\dfrac{d^{3}k_{1}}{(2\pi)^3}V\int_{}\dfrac{d^{3}k_{2}}{(2\pi)^3},
\label{16}
\end{equation}
where $\varrho=V^{-1}$ is the charge density, and $|J_{inc}|=\sqrt{(q_{1}q_{2})^{2}-m_{e}^{*^{4}}}/({Q_{1}Q_{2}V})$ is the incident particle current in the centre of mass frame.
After simplifications and by averaging over the polarizations of the incoming particles, and summing over the final ones, we get:
\small
\begin{eqnarray}
d\bar{\sigma_{n}}({e}^{+}{e}^{-}\rightarrow H^{+}H^{-})&=&\dfrac{1}{16\sqrt{(q_{1}q_{2})^2-m_{e}^{*^{4}}}}   \big|\overline{A_{\gamma}^{n} + A_{Z}^{n}} \big|^{2} \int_{}\dfrac{|\mathbf{k}_{1}|^{2}d|\mathbf{k}_{1}|d\Omega}{(2\pi)^2E_{H^{-}}}\int_{}\dfrac{d^{3}k_{2}}{ E_{H^{+}}} \nonumber \\  &\times & \delta^{4}(k_{1}+k_{2}-q_{1}-q_{2}-nk).
\label{17}
\end{eqnarray}
\normalsize
Consequently, the differential quantity has to be integrated over all four-momentum variables, and the integration over $d|\mathbf{k}_{1}|$ can performed by using the well known formula \cite{36} given by:
\begin{equation}
 \int d\mathbf y f(\mathbf y) \delta(g(\mathbf y))=\dfrac{f(\mathbf y)}{|g^{'}(\mathbf y)|_{g(\mathbf y)=0}}.
 \label{18}
\end{equation}
Finally, the expression of the differential cross section becomes as follows:
\small
\begin{eqnarray}
\dfrac{d\bar{\sigma_{n}}}{d\Omega}({e}^{+}{e}^{-}\rightarrow H^{+}H^{-})&=&\dfrac{1}{16\sqrt{(q_{1}q_{2})^2-m_{e}^{*^{4}}}}   \big|\overline{A_{\gamma}^{n} + A_{Z}^{n}} \big|^{2} \dfrac{2|\mathbf{k}_{1}|^{2}}{(2\pi)^2E_{H^{-}}}\dfrac{1}{g^{'}(|\mathbf{k}_{1}|)},
\label{19}
\end{eqnarray}
\normalsize
where the function $g^{'}(|\mathbf{k}_{1}|)$ is given by:
\begin{equation}
 g^{'}(|\mathbf{k}_{1}|)=\dfrac{2|\mathbf{k}_{1}|}{\sqrt{|\mathbf{k}_{1}|^{2}+M_{H^{-}}^{2}}}\Bigg( \dfrac{2e^{2}a^{2}}{\sqrt{s}} - n\omega -\sqrt{s} \Bigg).
 \label{20}
\end{equation}
The expression of the quantity $\big|\overline{A_{\gamma}^{n} + A_{Z}^{n}} \big|^{2}$ is given in the appendix.
\section{Results and discussion}
\begin{figure}[H]
  \centering
      \includegraphics[scale=0.7]{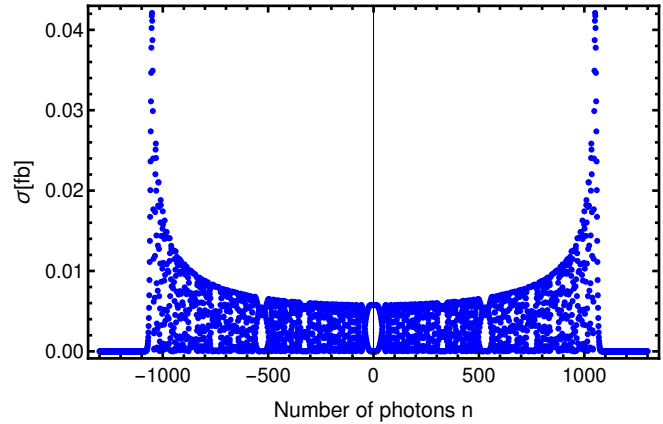}
  \caption{Laser-assisted partial total cross section (PTCS) as a function of the number of exchanged photons by taking the centre of mass energy and the charged Higgs mass as: $\sqrt{s}=500\,GeV$ ; $M_{H^{\pm}}=311\,GeV$. The laser field parameters are taken as $\varepsilon_{0}=10^{7}\,V.cm^{-1}$ and $\hbar\omega=1.17\,eV$.}
  \label{fig2}
\end{figure}

\begin{figure}[H]
  \centering
     \includegraphics[scale=0.55]{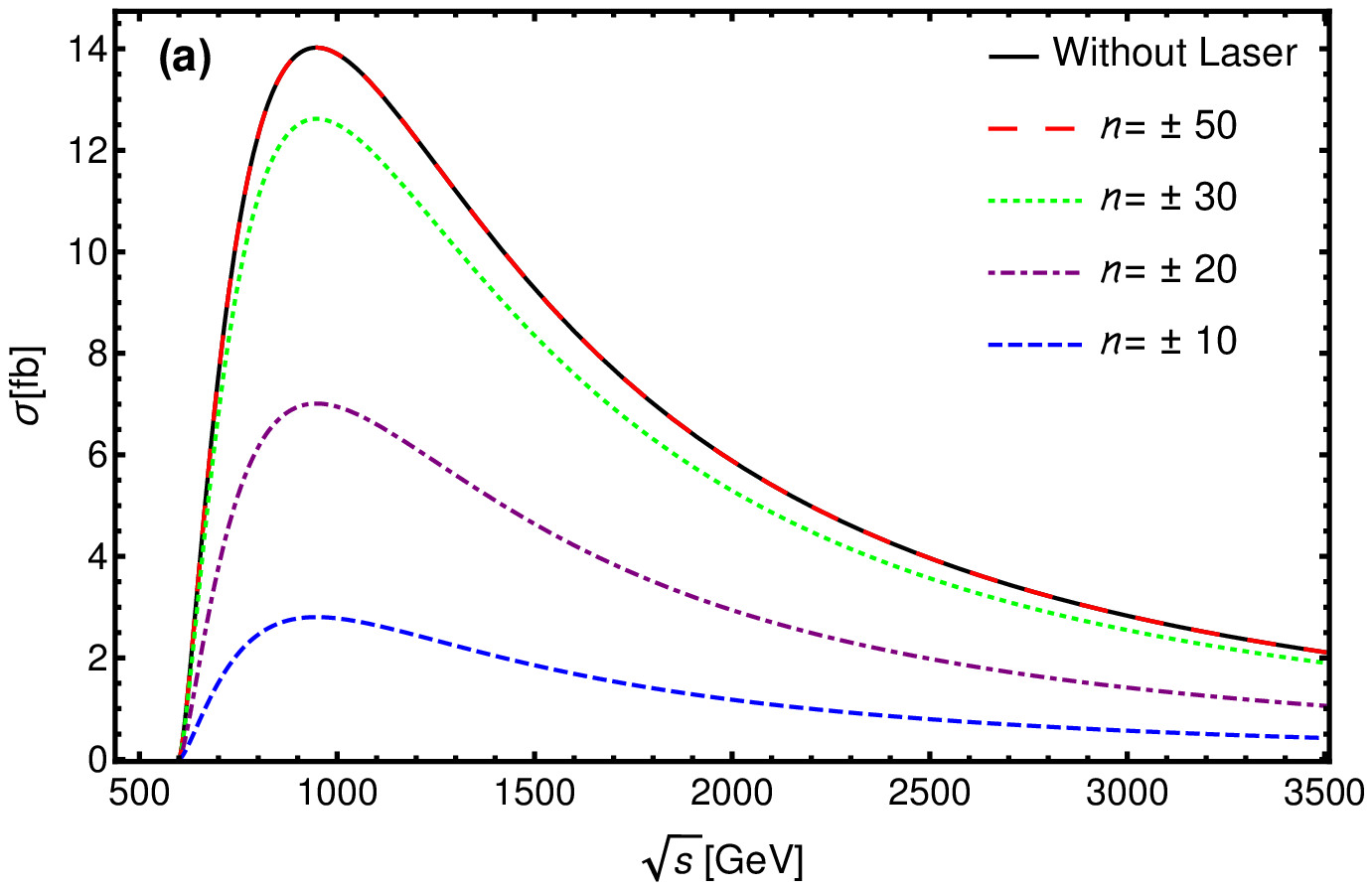}\hspace*{0.4cm}
      \includegraphics[scale=0.55]{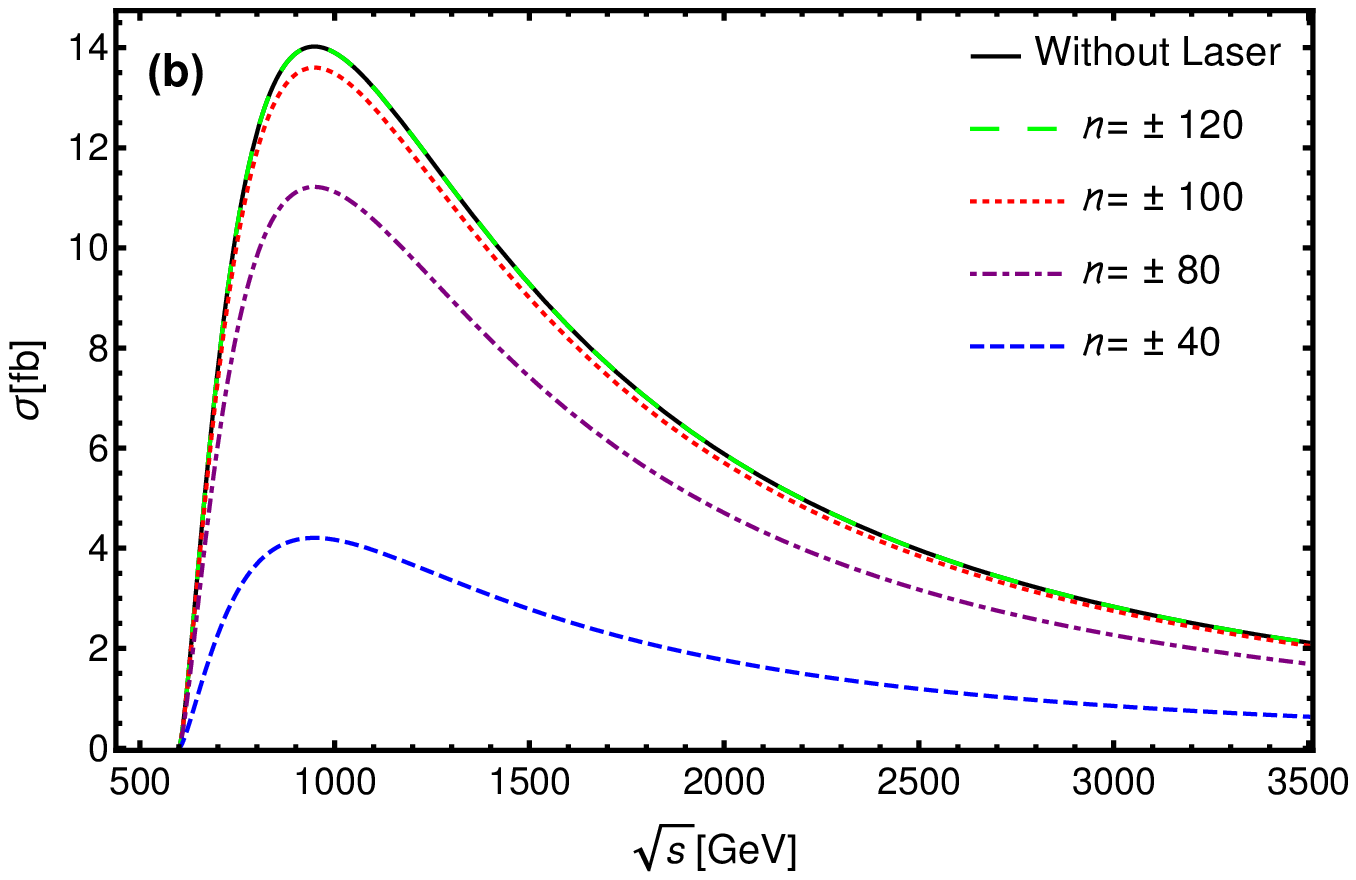}\par\vspace*{0.5cm}
      \includegraphics[scale=0.55]{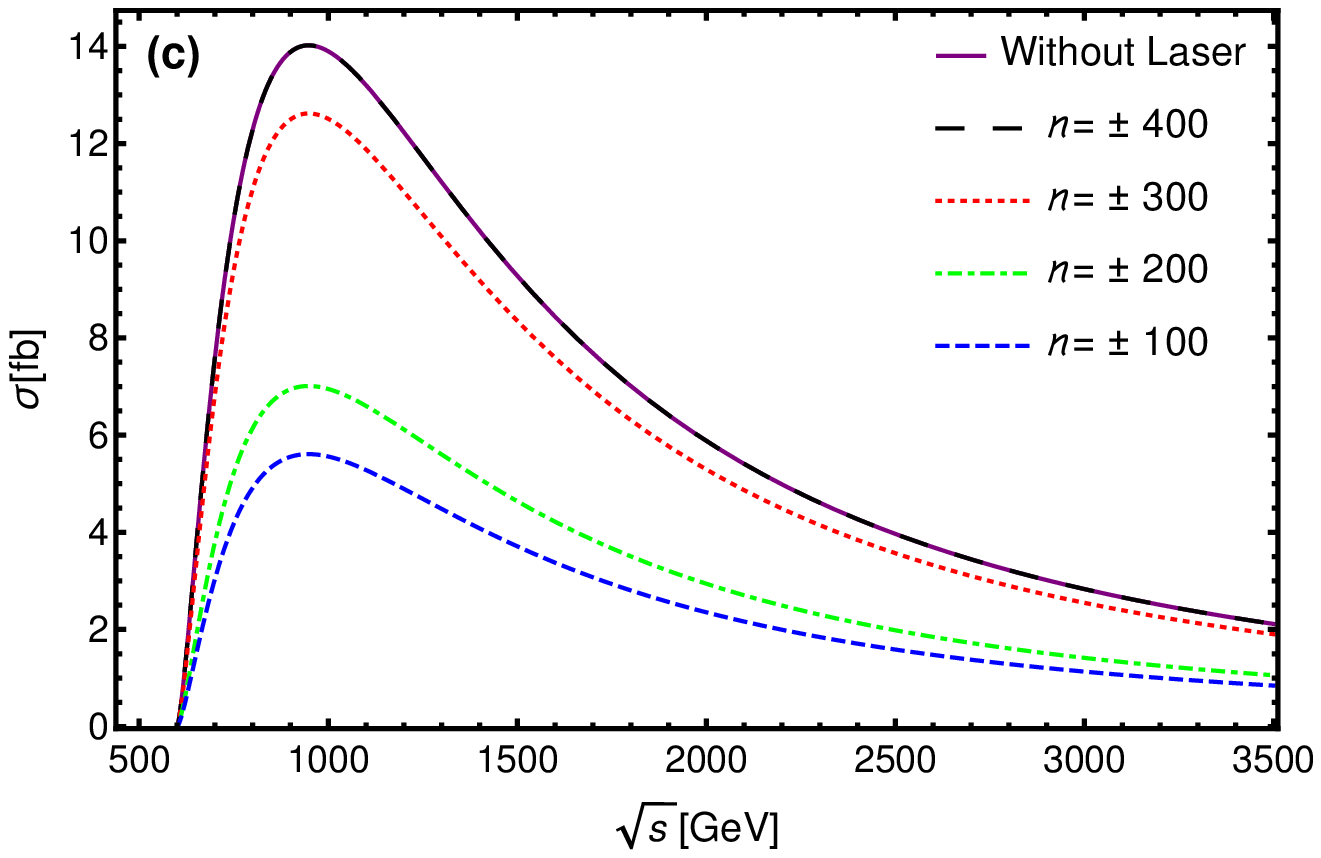}\hspace*{0.4cm}
      \includegraphics[scale=0.55]{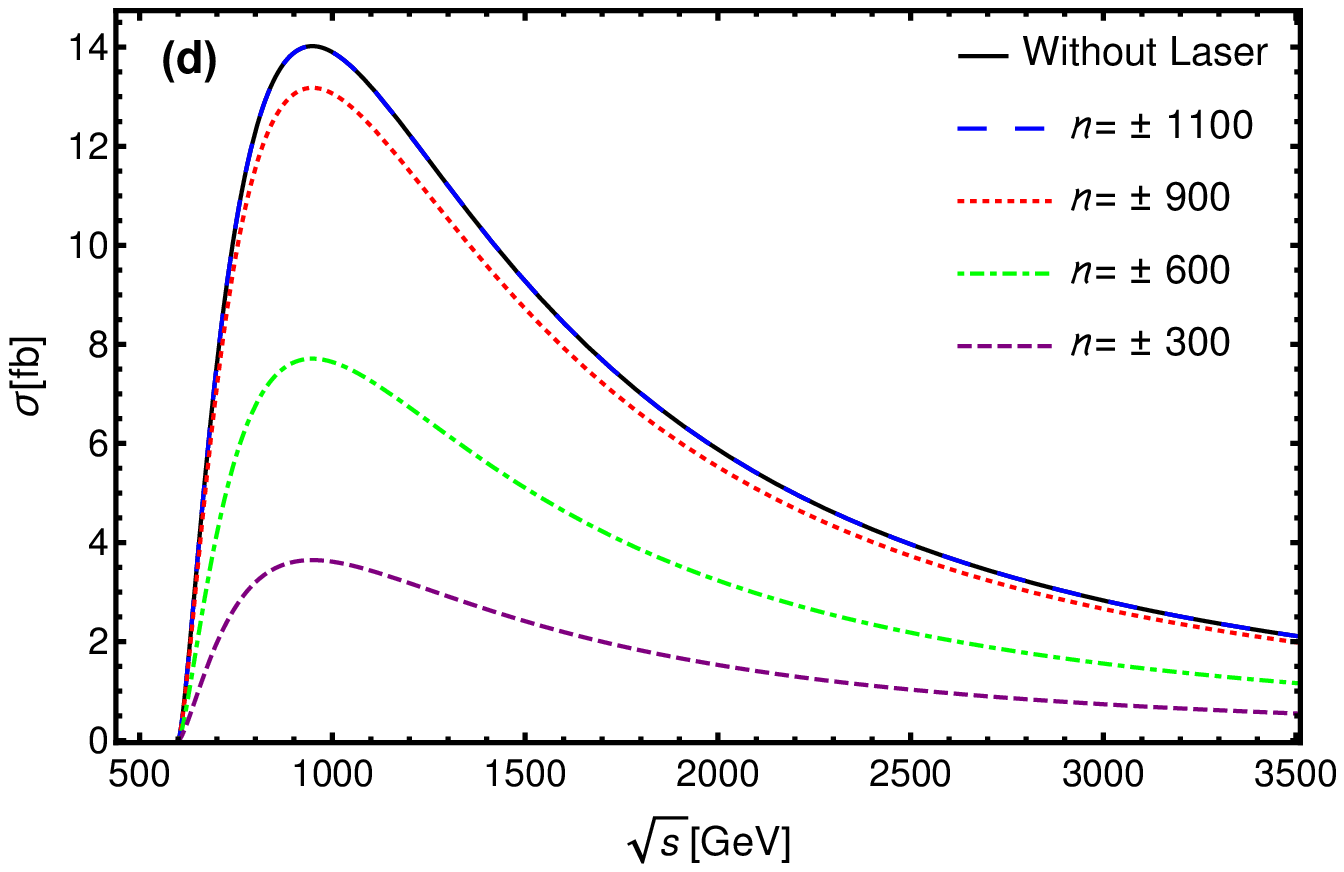}\par\vspace*{0.5cm}
        \caption{Dependence of the laser-assisted total cross section of the process ${e}^{+}{e}^{-}\rightarrow H^{+}H^{-}$ on the centre of mass energy for different exchanged photons number and by taking the charged Higgs mass as $M_{H^{\pm}}=311\,GeV$. The laser field strength and its frequency are chosen as: (a): $\varepsilon_{0}=10^{6}\,V.cm^{-1}$ , $\,\omega=2\, eV$ ; (b): $\varepsilon_{0}=10^{6}\,V.cm^{-1}$ , $\,\omega=1.17\, eV$ ; (c): $\varepsilon_{0}=10^{7}\,V.cm^{-1}$ , $\,\omega=2\, eV$ ; (d): $\varepsilon_{0}=10^{7}\,V.cm^{-1}$ , $\,\omega=1.17 \,eV$.}
        \label{fig3}
\end{figure}

In this section, the numerical results for the charged Higgs pair production in an $e^{+}e^{-}$ collider are presented and discussed in the centre of mass frame. 
It is well known that there exist two types of laser-matter interactions. The first type is laser-induced processes which can not occur outside the electromagnetic field. The second type occurs either in the absence or presence of the electromagnetic field with the same final products though the interaction properties are modified in its presence. The studied process can be included in the second type of laser-matter interaction. We have to mention that the polarization vector of the electromagnetic field is chosen to be parallel to the $Z$-axis. The total cross section is obtained from the differential cross section given by the equation (\ref{19}) by numerically integrating over the solid angle $d\Omega=\sin (\theta)d\theta d\phi$, where $\theta$ is the scattering angle. It should be noted that the laser-free production cross section of the process $e^{+}e^{-}\rightarrow H^{+}H^{-}$, at the lowest order, depends only on the charged  Higgs mass and the centre of mass energy \cite{37}. However, its corresponding laser-assisted cross section depends also on the laser field parameters. The current parameters of the SM are taken from PDG \cite{38} such that $m_{e} = 0.511\, MeV$, $m_{Z}=91.1875\,GeV$ and the Fermi coupling constant $G_{F}=1.166 3787 \times 10^{-5}\, GeV^{-2}$. The charged Higgs mass is chosen in such a way that it is consistent with dark matter constraints. Thus, $M_{H^{\pm}}=311\,GeV$ in figures \ref{fig2}, \ref{fig3} and \ref{fig5}(a).
The first step in this discussion concerns the behavior of the partial total cross section, which corresponds to each four-momentum conservation $\delta(p_{3}+p_{4}-q_{1}-q_{2}-nk)=0$, as a function of the number of exchanged photons.
\begin{table}[H]
\small
 \centering
\caption{\label{tab1}Laser-assisted Partial Total Cross Section (PTCS) as a function of the number of exchanged photons for different laser field strengths and frequencies. The centre of mass energy and the charged Higgs mass are chosen as: $\sqrt{s}=500\,GeV$ ; $M_{H^{\pm}}=311\,GeV$.}
\begin{tabular}{cccccccc}
\hline
  & PTCS[fb]  & &PTCS[fb] & & PTCS[fb] & & PTCS[fb] \\
  &  $\varepsilon_{0}=10^{6}V/cm $ & & $\varepsilon_{0}=10^{6}V/cm $& &  $\varepsilon_{0}=10^{7}V/cm $& &  $\varepsilon_{0}=10^{7}V/cm $\\
 n &  $\omega=2eV $ & n &$\omega=1.17eV $& n & $\omega=2eV  $ & n & $\omega=1.17eV  $\\
 \hline
 $-70$ & $ 0 $ & $-140$ & $ 0 $ & $-500$ & $ 0 $ & $-1300$ & $ 0 $\\
 $-50$ & $ 0 $ & $-120$ & $ 0 $ & $-400$ & $ 0 $ & $-1100$ & $ 0 $\\  
   $-30$ & $ 0.00239246 $ & $-100$ & $ 0.130679 $ & $-300$ & $ 0.00405759 $ & $-900$ & $ 0.00017812 $\\
    $-20$ & $ 0.184772 $ & $-80$ & $ 0.0182863 $ & $-200$ & $ 0.0124957 $ & $-600$ & $ 0.0019426 $ \\
   $-10$ & $ 0.0775327 $ & $-40$ & $ 0.0570623 $ & $-100$ & $ 0.00106369 $ & $-300$ & $ 0.000581094 $\\
    $0$ & $ 0.0632618 $ & $0$ & $ 0.000400618 $ & $0$ & $ 0.012883 $ & $0$ & $ 0.00575724 $\\
    $10$ & $ 0.0775327  $ & $40$ & $ 0.0570623 $  & $100$ & $ 0.00106369 $ & $300$ & $ 0.000581094 $\\
    $20$ & $ 0.184772 $ & $80$ & $ 0.0182863 $& $200$ & $ 0.0124957 $ & $600$ & $  0.0019426 $\\
    $30$ & $ 0.00239246 $ & $100$ & $ 0.130679 $ & $300$ & $ 0.00405759 $ & $900$ & $ 0.00017812 $\\
   $50$ & $ 0 $  & $120$ & $ 0 $ & $400$ & $ 0 $ & $1100$ & $ 0 $ \\
   $70$ & $ 0 $  & $140$ & $ 0 $ & $500$ & $ 0 $ & $1300$ & $ 0 $ \\
     \hline
\end{tabular}
\normalsize
\end{table}
Table \ref{tab1} represents the PTCS of the charged Higgs pair production process versus the number of exchanged photons in the presence of a circularly polarized electromagnetic field. This dependence is illustrated for different photons number, laser field strengths and frequencies. Regardless of the exchanged photons number, we observe that the partial total cross section which corresponds to a number $n$ of photons is the same for $-n$. In addition, as far as  the number $n$ is raised, the PTCS decreases until it becomes zero. The first number of exchanged photons, for which the PTCS is equal to zero is called cutoff. It depends on the other laser field parameters, i.e, the laser field strength and its frequency. Therefore , the PTCS presents two symmetric cutoffs for each laser field strength and frequency. By comparing the PTCS's, we observe that the cutoff number increases either by fixing the laser field frequency and increasing its strength or by decreasing the laser field frequency and fixing its strength. To display clearly these results, we have plotted in figure \ref{fig2} the PTCS which corresponds to $\varepsilon_{0}=10^{7}\,V.cm^{-1}$ and $\hbar\omega=0.117\,eV$. This figure confirms the symmetric aspect of the laser-assisted PTCS. Therefore, the PTCS of emission of photons has the same order as that corresponds to absorption.
No, let's move on to discuss the variation of the total cross section of the charged Higgs pair production on the centre of mass energy. 
Figure \ref{fig3} displays the laser-assisted total cross section versus the centre of mass energy for different number of exchanged photons and for different laser field strengths and frequencies which are already considered in Table \ref{tab1}. By beginning with the laser-free total cross section, we notice that the total cross section raises from $\sqrt{s}=589\, GeV$ until it reaches its maximum at $\sqrt{s}=944\,GeV$, then it begins to decrease as much as the centre of mass energy increases \cite{39}. By applying the circularly polarized laser field, the total cross section maintains the same general aspects, yet its order of magnitude decreases and depends on the laser parameters. For instance, in figure \ref{fig3}(a), in which  $\varepsilon_{0}=10^{6}\,V.cm^{-1}$ and $\,\omega=2 \,eV$, the total cross section increases by increasing the number of exchanged photons. For example, its maximum values are $\sigma=2.9\,fb$, $\sigma=7.06\,fb$ and $\sigma=12.68\,fb$ for $n=\pm 10$, $n=\pm 20$ and $n=\pm 30$, respectively. Moreover, when we reach $n=\pm 50$, the total cross section will be equal to its corresponding laser-free total cross section in all centre of mass energies with a maximum value $\sigma_{max}=14.02\,fb$. The same behavior happened in Figure \ref{fig3}(b), \ref{fig3}(c) and \ref{fig3}(d) as the total cross section raises as long as the number $n$ increases until $n=\pm$ cutoff. This summation over the cutoff number is called sum-rule, and it was elaborated by Kroll and Watson \cite{40}. It states that the summed laser-assisted partial cross section from $-$cutoff to $+$cutoff number will be equal to the laser-free total cross section.
Beyond $n=\pm$ cutoff, no photons will be exchanged between the laser field and the incident particles. Therefore, the laser-assisted total cross section is always equal to its corresponding laser free-total cross section. Another important point to be discussed, here, is that the intensity of the laser field and its frequency have a great impact on the total cross section. Moreover, the number of photons required to reach the well known sum-rule increases by either increasing the strength of the laser field or by decreasing its frequency. Furthermore, by comparing figures \ref{fig3}(c) and \ref{fig3}(d) where $\varepsilon_{0}=10^{7}\,V.cm^{-1}$, we remark that, for the same number of exchanged photons ( e.g., $n=\pm 300$), $\sigma=12.4\,fb$ for $\,\omega=2\, eV$ and $\sigma=2.80\,fb$ for $\,\omega=1.17\, eV$. In figure \ref{fig3}, the charged Higgs mass is chosen as $M_{H^{\pm}}=311\,GeV$, which is consistent with the IHDM constraints. Therefore, it is worthy and important to analyze the dependence of the total cross section on this free-parameter.
\begin{figure}[h!]
  \centering
      \includegraphics[scale=0.45]{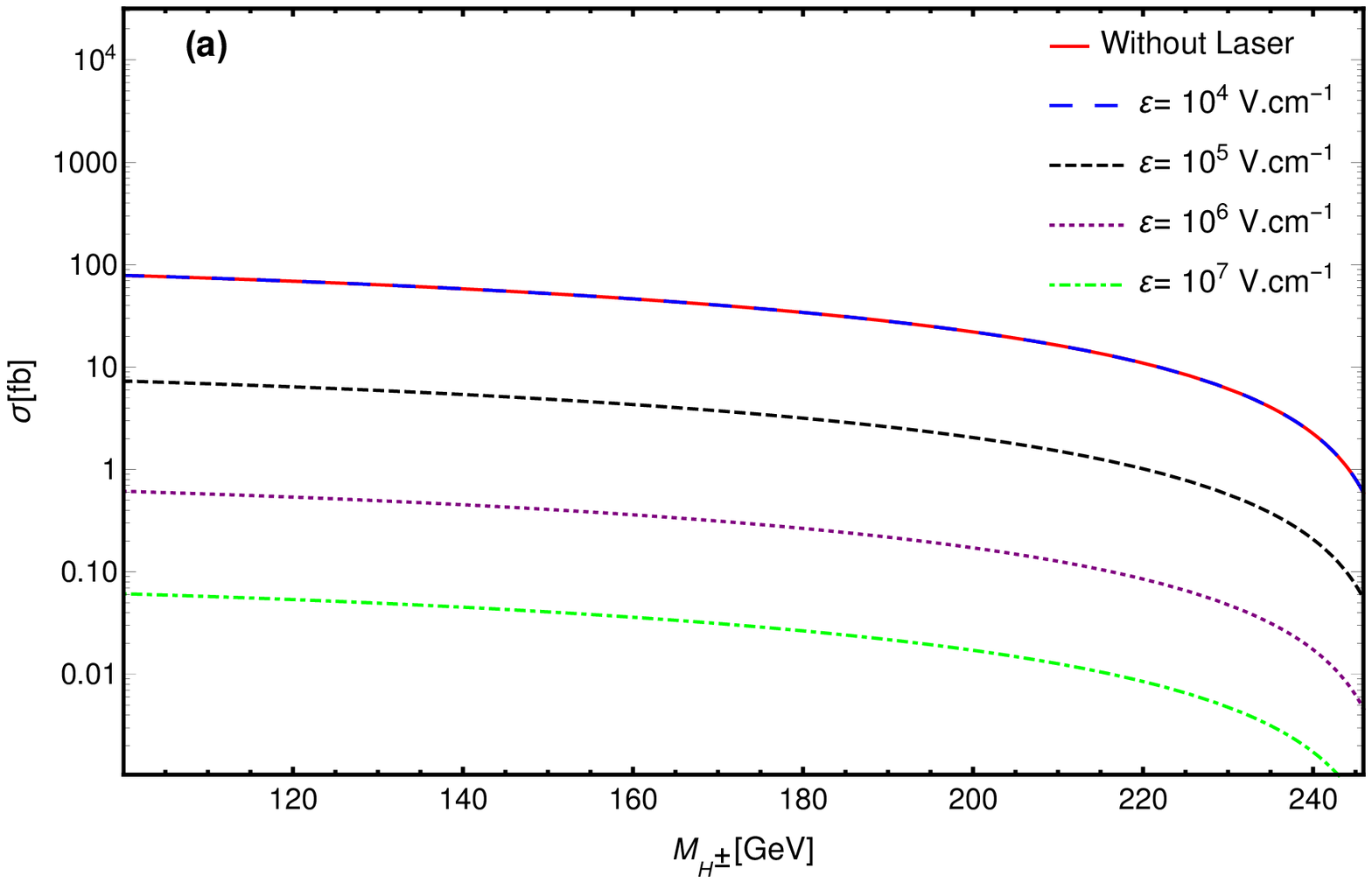}\hspace*{0.4cm}
      \includegraphics[scale=0.45]{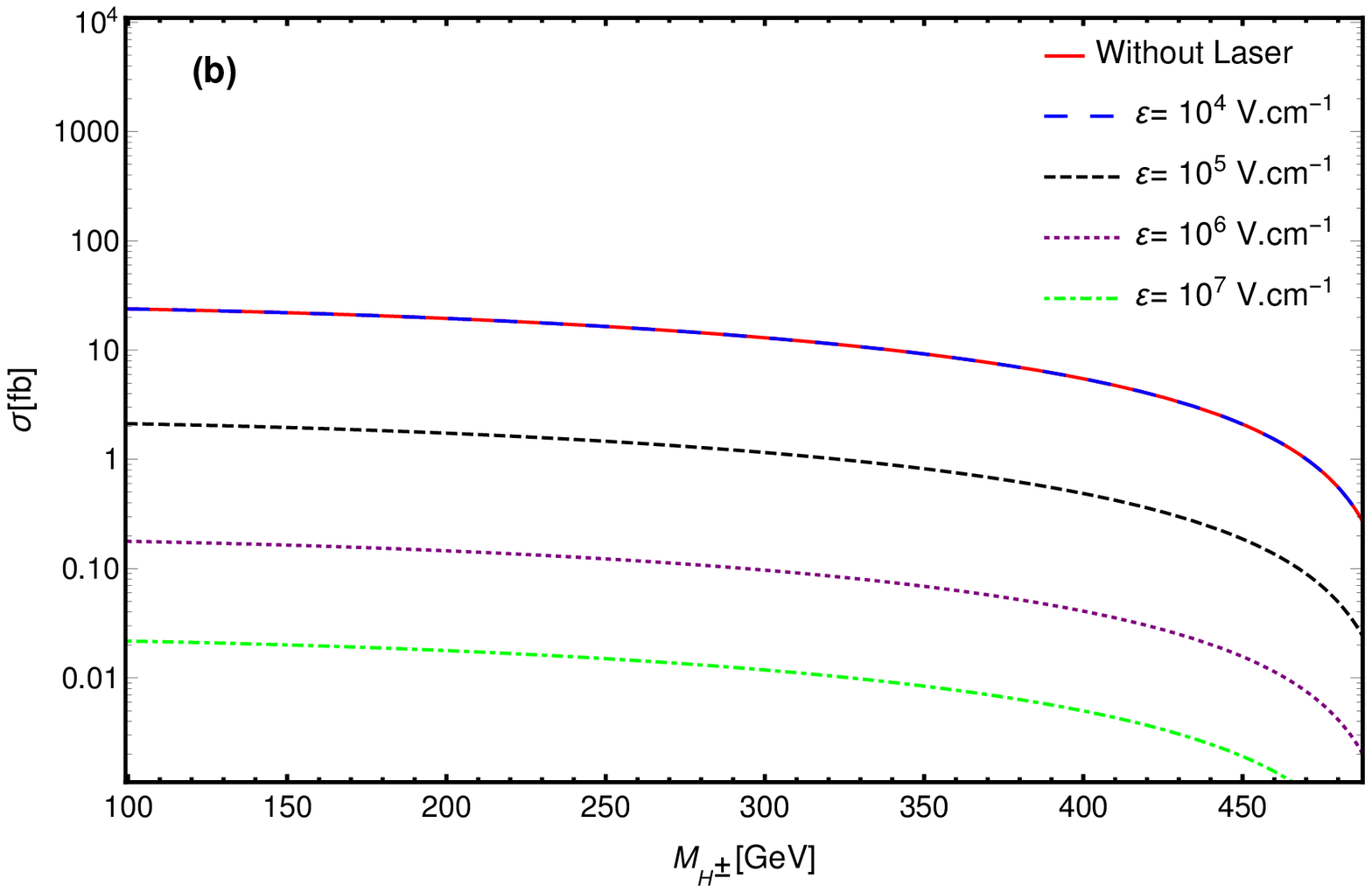}\par\vspace*{0.5cm}
      \includegraphics[scale=0.45]{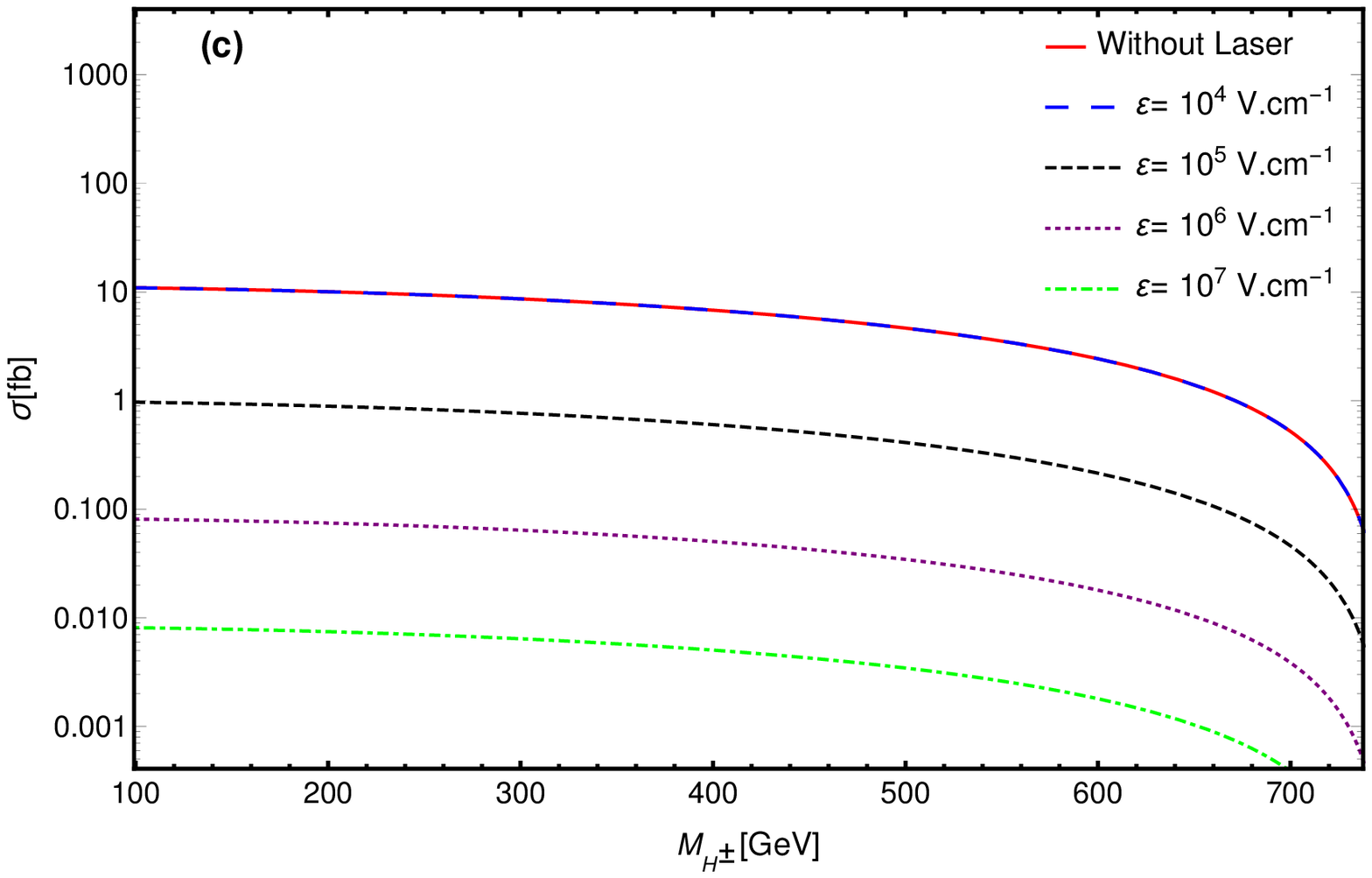}\hspace*{0.4cm}
      \includegraphics[scale=0.50]{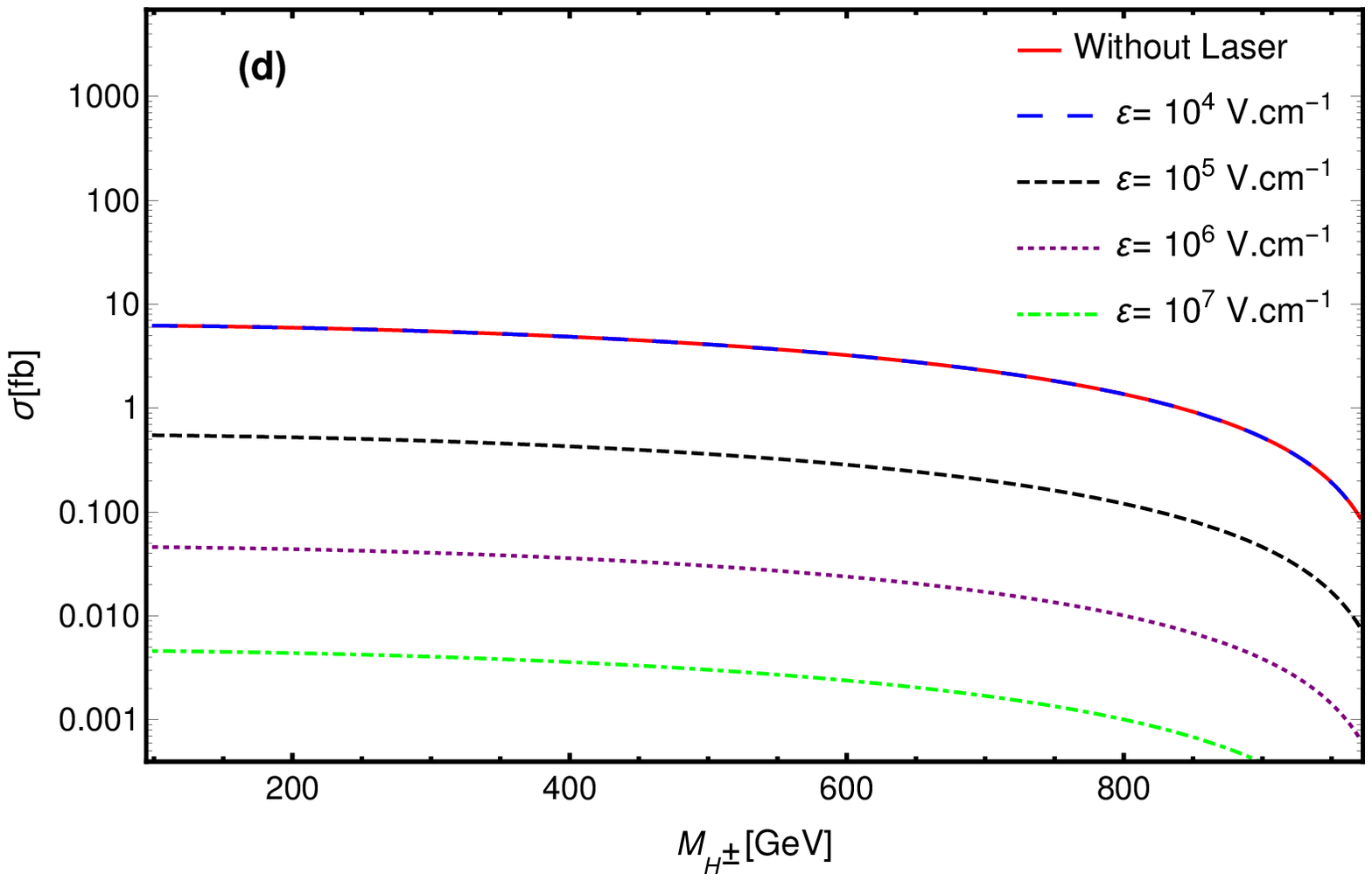}\par\vspace*{0.5cm}
        \caption{Variation of the laser-assisted total cross section of the process ${e}^{+}{e}^{-}\rightarrow H^{+}H^{-}$ as a function of the charged Higgs-boson mass and centre of mass energies by choosing the laser field frequency and the number of exchanged photons such that: $\omega=0.117\,eV$ , $n=\pm 900$. The centre of mass energy is taken as:(a): $\sqrt{s}=500\,GeV$ ; (b): $\sqrt{s}=1000\,GeV$ ; (c): $\sqrt{s}=1500\,GeV$ ; (d): $\sqrt{s}=2000\,GeV$.}
        \label{fig4}
\end{figure}
Figure \ref{fig4} illustrates the dependence of the laser-assisted total cross section on the mass of the produced pair of charged Higgs for different strengths of the electromagnetic field and for four typical centre of mass energies
and by fixing the number of exchanged photons as $n=\pm 900$. 
Let's begin by analyzing the behavior of the laser-free cross section. We remark that its order of magnitude depends on the centre of mass energy and decreases when $m_{H^\pm} \to \sqrt s/2$ (threshold effect). For instance, at low charged Higgs mass, the total cross section has the order of $100\, fb$, $25\, fb$, $10 \,fb$ and $6 \,fb$ for $\sqrt{s}=500\,GeV$, $\sqrt{s}=1000\,GeV$, $\sqrt{s}=1500\,GeV$ and $\sqrt{s}=2000\,GeV$, respectively. In addition and as it is expected from previous results, the laser-free cross section gets higher for low values of charged Higgs mass, then it slowly decreases as long as $M_{H^{\pm}}$ increases. This means that the cross-section values are inversely related to the mass of the charged Higgs.
According to Figure \ref{fig4}, for $\omega=0.117 \,eV$, the electromagnetic field does not affect the integrated laser-assisted cross section at low strengths $\varepsilon_{0}\leq 10^{4}\,V.cm^{-1}$. This is why the laser-assisted cross section is equal to its corresponding laser-free cross section for $\varepsilon_{0}= 10^{4}\,V.cm^{-1}$. Moreover, we observe that, regardless of the centre of mass energy, the order of magnitude of the total laser-assisted cross section decreases as much as the laser field strength increases. For example, in figure \ref{fig4} (a) and for $M_{H^{\pm}}=120\,GeV$, $\sigma=70.60\,fb$, $\sigma=6.47\,fb$, $\sigma=0.59\,fb$ and  $\sigma=0.057\,fb$ for the laser strengths $\varepsilon_{0}= 10^{4}\,V.cm^{-1}$, $\varepsilon_{0}= 10^{5}\,V.cm^{-1}$, $\varepsilon_{0}= 10^{6}\,V.cm^{-1}$ and $\varepsilon_{0}= 10^{7}\,V.cm^{-1}$, respectively.
Another important observation is that the laser-assisted total cross section falls down rapidly for high laser field strengths as compared to low strengths. This means that the threshold effect, which due to the phase space suppression, is shifted to $m_{H^\pm} > \sqrt s/2$ as much as we increase the laser field strength.
Given that the frequency of the laser field is also a crucial parameter, we have made the same analysis for different laser frequencies.
\begin{figure}[H]
  \centering
      \includegraphics[scale=0.50]{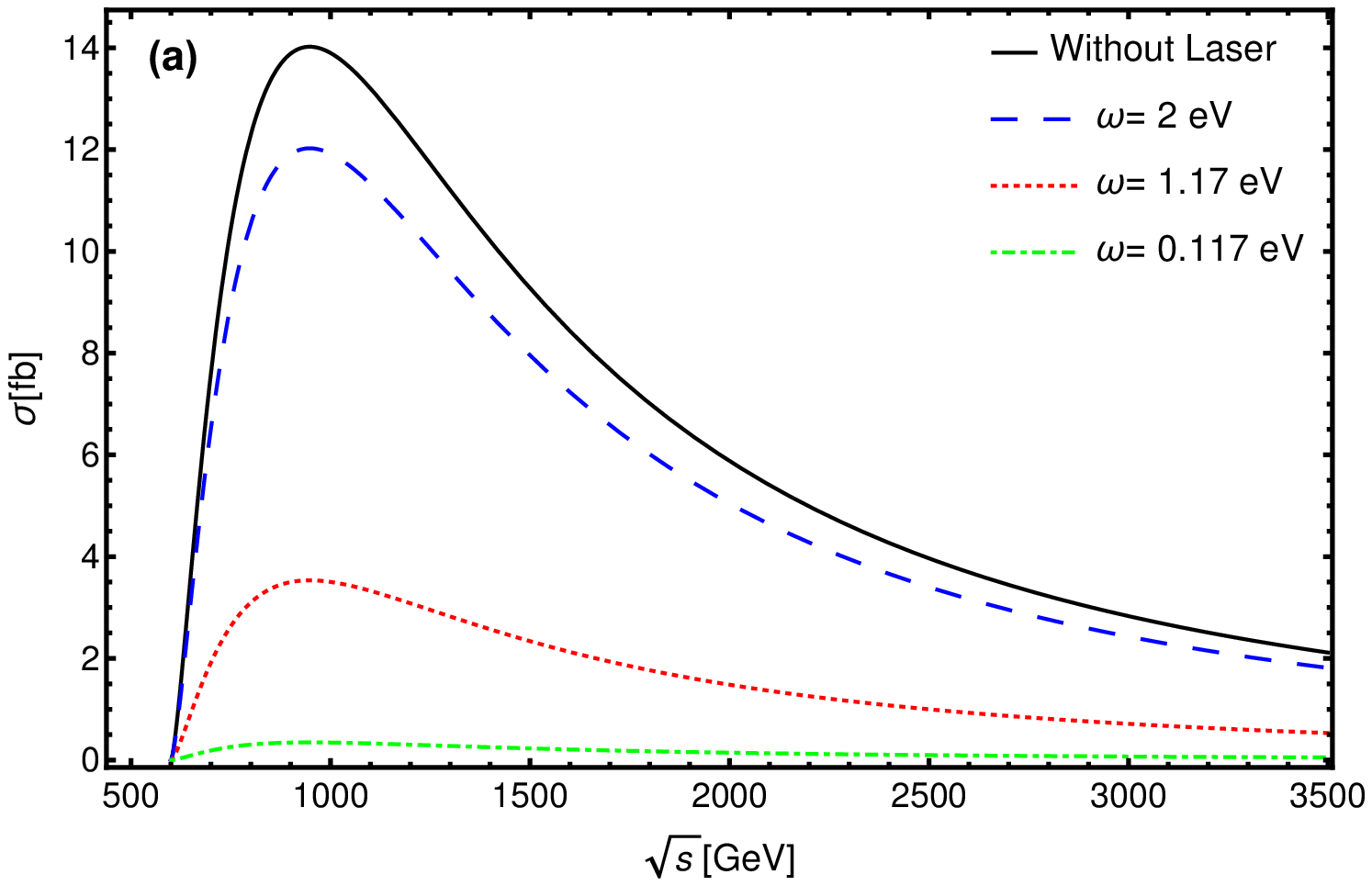}\hspace*{0.4cm}
      \includegraphics[scale=0.50]{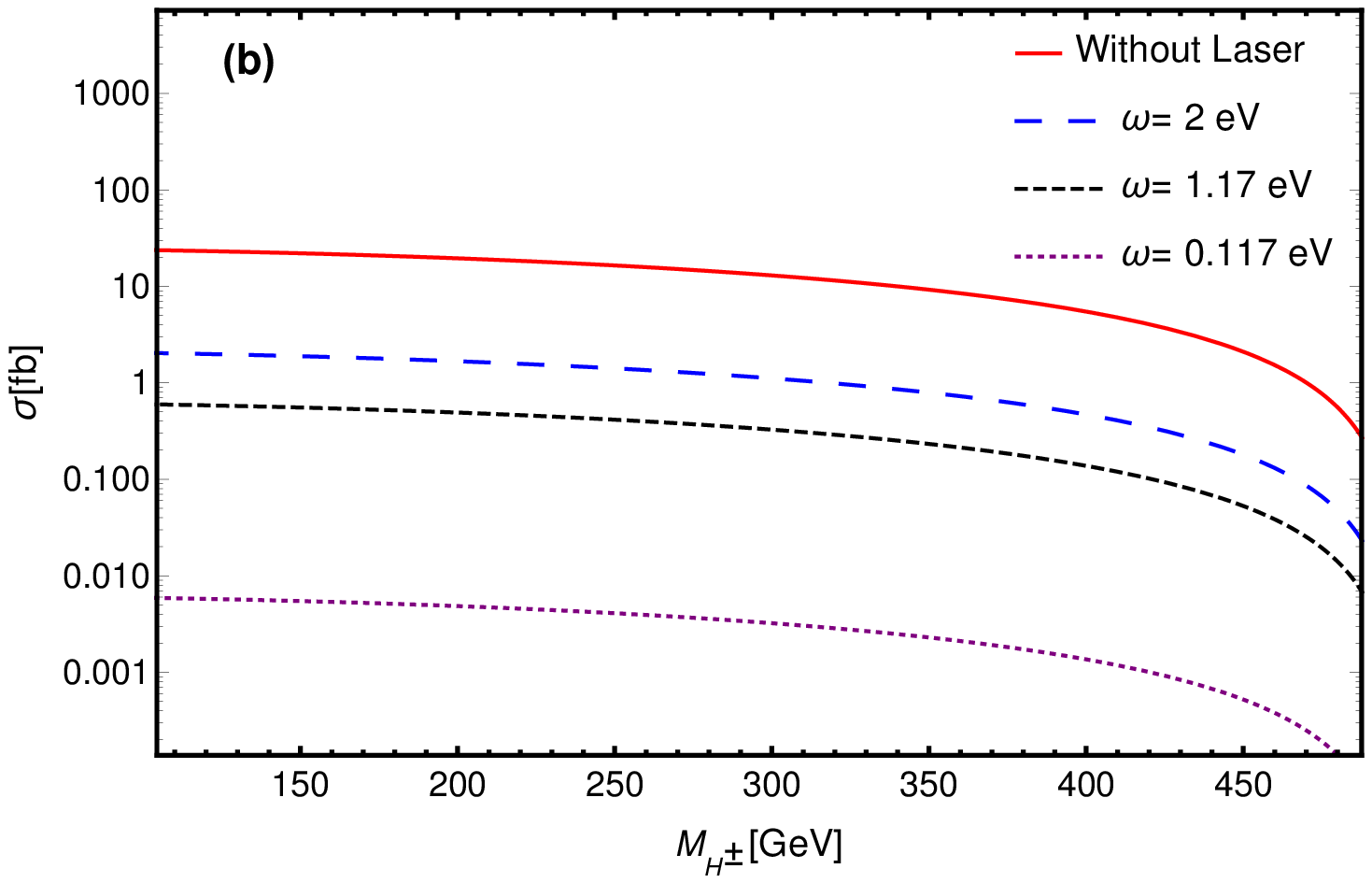}\par\vspace*{0.5cm}
        \caption{Laser-assisted total cross section of the process ${e}^{+}{e}^{-}\rightarrow H^{+}H^{-}$ for $n=\pm 900$ and $\varepsilon_{0}=10^{7}\,V.cm^{-1}$. (a):  $M_{H^{\pm}}=311\,GeV$ ; (b): $\sqrt{s}=1000\,GeV$.}
        \label{fig5}
\end{figure}
Figure \ref{fig5} shows the dependence of the laser-assisted total cross section on the centre of mass energy and the charged Higgs mass for different frequencies. The number of exchanged photons is arbitrary chosen as $n=\pm 900$. As we can see from figure \ref{fig5}(a), the order of magnitude of the cross section decreases by decreasing the frequency of the laser field. The cross section as a function of  $M_{H^{\pm}}$ in figure \ref{fig5}(b) behaves in the same manner as its order of magnitude raises when $\omega$ increases. This is due to the fact that the number of photons, required to reach the well know sum-rule, depends on the frequency $\omega$. More precisely, this number increases as much as the frequency of the electromagnetic field increases.
\section{Conclusion}
In the present paper, we have investigated the pair of charged Higgs production in IHDM in the presence of a circularly polarized electromagnetic field. First, we have illustrated the dependence of the partial total cross section on the number of exchanged photons, and we have shown that the cutoff number depends on the laser field strength and its frequency as it increases by increasing the laser field strength or by decreasing the laser frequency. Secondly,  we have found that the order of magnitude of the total cross section decreases by increasing the centre of mass energy. However, this increasing process is slow for low centre of mass energy in comparison to high $\sqrt{s}$. Then we have indicated that the laser-assisted cross section as a function of $M_{H^{\pm}}$ falls down rapidly as far as we increase the laser field strength or by decreasing its frequency.
\section*{Appendix}
We have given in this appendix, the expression of the term $\big|\overline{A_{\gamma}^{n} + A_{Z}^{n}} \big|^{2}$ that appears in equation (\ref{19}).
\tiny
\begin{eqnarray}
 \big|\overline{A_{\gamma}^{n} + A_{Z}^{n}} \big|^{2}&=&\nonumber\dfrac{1}{4}\sum_{n=-\infty}^{+\infty}\sum_{s}\big| A_{\gamma}^{n} + A_{Z}^{n} \big|^{2}=\frac{1}{4}\sum_{n=-\infty}^{+\infty}\Bigg\lbrace\dfrac{e^{4}}{(q_{1}+q_{2}+nk)^{4}} Tr\Bigg[(\slashed p_{1}-m_{e})(k_{2}^{\mu}-k_{1}^{\mu})\\ \nonumber &\times &\Big[ \chi^{0}_{\mu}\,J_{n}(z)e^{-in\phi _{0}}(z)+\chi^{1}_{\mu}\,\,\frac{1}{2}\Big(J_{n+1}(z)e^{-i(n+1)\phi _{0}} + J_{n-1}(z)e^{-i(n-1)\phi _{0}}\Big) \\ &+ & \nonumber\chi^{2}_{\mu}\,\frac{1}{2\, i}\Big(J_{n+1}(z)e^{-i(n+1)\phi _{0}}-J_{n-1}(z)e^{-i(n-1)\phi _{0}}\Big)\Big](\slashed p_{2}+m_{e})(k_{2}^{\nu}-k_{1}^{\nu})\\ \nonumber &\times &  \Big[ \chi^{0}_{\nu}\,J^{*}_{n}(z)e^{+in\phi _{0}}(z)+\chi^{1}_{\nu}\,\,\frac{1}{2}\Big(J^{*}_{n+1}(z)e^{+i(n+1)\phi _{0}} + J^{*}_{n-1}(z)e^{+i(n-1)\phi _{0}}\Big) \\ &- & \nonumber\chi^{2}_{\nu}\,\frac{1}{2\, i}\Big(J^{*}_{n+1}(z)e^{+i(n+1)\phi _{0}}-J^{*}_{n-1}(z)e^{+i(n-1)\phi _{0}}\Big)\Big]\Bigg]
+\left(\dfrac{e}{2C_{W}S_{W}}\right)^{2} \left( \dfrac{g }{C_{W}}\Big(\frac{1}{2}-S_{W}^{2}\Big)\right)^{2}\\ \nonumber &\times &  \left(\frac{1}{(q_{1}+q_{2}+nk)^{2}-M_{Z}^{2}}\right)^{2} Tr\Bigg[(\slashed p_{1}-m_{e})(k_{2}^{\mu}-k_{1}^{\mu})\Big[ \lambda^{0}_{\mu}\,J_{n}(z)e^{-in\phi _{0}}(z)\\ \nonumber &+ &\lambda^{1}_{\mu}\,\,\frac{1}{2}\Big(J_{n+1}(z)e^{-i(n+1)\phi _{0}} + J_{n-1}(z)e^{-i(n-1)\phi _{0}}\Big)+\lambda^{2}_{\mu}\,\frac{1}{2\, i}\Big(J_{n+1}(z)e^{-i(n+1)\phi _{0}}-J_{n-1}(z)e^{-i(n-1)\phi _{0}}\Big)\Big]\\ \nonumber &\times & (\slashed p_{2}+m_{e})(k_{2}^{\nu}-k_{1}^{\nu}) \Big[ \lambda^{0}_{\nu}\,J^{*}_{n}(z)e^{+in\phi _{0}}(z)+ \lambda^{1}_{\nu}\,\,\frac{1}{2}\Big(J^{*}_{n+1}(z)e^{+i(n+1)\phi _{0}} + J^{*}_{n-1}(z)e^{+i(n-1)\phi _{0}}\Big) \\ &- & \nonumber\lambda^{2}_{\nu}\,\frac{1}{2\, i}\Big(J^{*}_{n+1}(z)e^{+i(n+1)\phi _{0}}-J^{*}_{n-1}(z)e^{+i(n-1)\phi _{0}}\Big)\Big]\Bigg]
+ \dfrac{e^{2}}{(q_{1}+q_{2}+nk)^{2}}\left(\dfrac{e}{2C_{W}S_{W}}\right)\\ \nonumber &\times & \left( \dfrac{g }{C_{W}}\Big(\frac{1}{2}-S_{W}^{2}\Big)\right)  \frac{1}{(q_{1}+q_{2}+nk)^{2}-M_{Z}^{2}} Tr\Bigg[(\slashed p_{1}-m_{e})(k_{2}^{\mu}-k_{1}^{\mu})\Big[ \chi^{0}_{\mu}\,J_{n}(z)e^{-in\phi _{0}}(z) \\ &+ &\nonumber\chi^{1}_{\mu}\,\,\frac{1}{2}\Big(J_{n+1}(z)e^{-i(n+1)\phi _{0}} + J_{n-1}(z)e^{-i(n-1)\phi _{0}}\Big)+ \chi^{2}_{\mu}\,\frac{1}{2\, i}\Big(J_{n+1}(z)e^{-i(n+1)\phi _{0}}-J_{n-1}(z)e^{-i(n-1)\phi _{0}}\Big)\Big]\\ \nonumber &\times & (\slashed p_{2}+m_{e})(k_{2}^{\nu}-k_{1}^{\nu}) \Big[ \lambda^{0}_{\nu}\,J^{*}_{n}(z)e^{+in\phi _{0}}(z)+\lambda^{1}_{\nu}\,\,\frac{1}{2}\Big(J^{*}_{n+1}(z)e^{+i(n+1)\phi _{0}} + J^{*}_{n-1}(z)e^{+i(n-1)\phi _{0}}\Big)\\ \nonumber &- & \lambda^{2}_{\nu}\,\frac{1}{2\, i}\Big(J^{*}_{n+1}(z)e^{+i(n+1)\phi _{0}}-J^{*}_{n-1}(z)e^{+i(n-1)\phi _{0}}\Big)\Big]\Bigg]
+\dfrac{e^{2}}{(q_{1}+q_{2}+nk)^{2}}\left(\dfrac{e}{2C_{W}S_{W}}\right)\\ \nonumber &\times & \left( \dfrac{g }{C_{W}}\Big(\frac{1}{2}-S_{W}^{2}\Big)\right)  \frac{1}{(q_{1}+q_{2}+nk)^{2}-M_{Z}^{2}} Tr\Bigg[(\slashed p_{1}-m_{e})(k_{2}^{\mu}-k_{1}^{\mu})\\ \nonumber &\times & \Big[ \lambda^{0}_{\mu}\,J_{n}(z)e^{-in\phi _{0}}(z)+\lambda^{1}_{\mu}\,\,\frac{1}{2}\Big(J_{n+1}(z)e^{-i(n+1)\phi _{0}} + J_{n-1}(z)e^{-i(n-1)\phi _{0}}\Big)\\ \nonumber &+ & \lambda^{2}_{\mu}\,\frac{1}{2\, i}\Big(J_{n+1}(z)e^{-i(n+1)\phi _{0}}-J_{n-1}(z)e^{-i(n-1)\phi _{0}}\Big)\Big] (\slashed p_{2}+m_{e})(k_{2}^{\nu}-k_{1}^{\nu}) \Big[ \chi^{0}_{\nu}\,J^{*}_{n}(z)e^{+in\phi _{0}}(z)\\ &+ & \nonumber\chi^{1}_{\nu}\,\,\frac{1}{2}\Big(J^{*}_{n+1}(z)e^{+i(n+1)\phi _{0}} + J^{*}_{n-1}(z)e^{+i(n-1)\phi _{0}}\Big) -\chi^{2}_{\nu}\,\frac{1}{2\, i}\Big(J^{*}_{n+1}(z)e^{+i(n+1)\phi _{0}}-J^{*}_{n-1}(z)e^{+i(n-1)\phi _{0}}\Big)\Big]\Bigg]\Bigg\rbrace
\end{eqnarray}


\normalsize
\begin{thebibliography}{99}
\bibitem{1} S.W. Bahk, et al., Optics Lett.\textbf{29} (2004) 2837, https://doi.org/10.1364/OL.29.002837.

\bibitem{2} G.~A.~Mourou, T.~Tajima and S.~V.~Bulanov,
%``Optics in the relativistic regime,''
Rev. Mod. Phys. \textbf{78} (2006) 309, 
doi:10.1103/RevModPhys.78.309.

\bibitem{3}C.~Muller, K.~Z.~Hatsagortsyan and C.~H.~Keitel,
%``Muon pair creation from positronium in a circularly polarized laser field,''
Phys. Rev. D \textbf{74} (2006) 074017,
doi:10.1103/PhysRevD.74.074017,
[arXiv:physics/0602106 [physics.atom-ph]].

\bibitem{4}M.~Jakha, S.~Mouslih, S.~Taj and B.~Manaut,
%``Laser effect on the final products of Z-boson decay,''
Laser Phys. Lett. \textbf{18} (2021) 016002,
doi:10.1088/1612-202X/abd17d
[arXiv:2010.14401 [hep-ph]].

\bibitem{5} S.~P.~Roshchupkin, V.~V.~Dubov and A.~Dubov,
%``Resonant effects in the spontaneous bremsstrahlung process of ultrarelativistic electrons in the fields of a nucleus and a pulsed light wave,''
Laser Phys. Lett. \textbf{18} (2021) 045301,
doi:10.1088/1612-202X/abeb21,
[arXiv:2004.02247 [quant-ph]].

\bibitem{6}A.~H.~Liu, S.~M.~Li and J.~Berakdar,
%``Laser-assisted muon decay,''
Phys. Rev. Lett. \textbf{98} (2007) 251803,
doi:10.1103/PhysRevLett.98.251803.

\bibitem{7}M. Ouali, M. Ouhammou, S. Taj, and B. Manaut, 
\textit{" $Z$-boson production via the weak process ${e}^{+}{e}^{-}\rightarrow\mu^{+} \mu^{-}$ in the presence of a circularly polarized laser field "},
[arXiv:2105.14854 [hep-ph]].

\bibitem{8}M.~Ouhammou, \textit{et al.},
%``Higgs-strahlung boson production in the presence of a circularly polarized laser field,''
Laser Phys. Lett. \textbf{18} (2021) 076002,
doi:10.1088/1612-202X/ac0919
[arXiv:2104.11155 [hep-ph]].

\bibitem{9} M. Ouhammou, M. Ouali, S. Taj, and B. Manaut, \textit{" Laser-assisted neutral Higgs-boson pair production in Inert Higgs Doublet Model (IHDM) "}, [arXiv:2108.04937 [hep-ph]].

\bibitem{10} S.~Mouslih, M.~Jakha, S.~Taj, B.~Manaut and E.~Siher,
%``Laser-assisted pion decay,''
Phys. Rev. D \textbf{102} (2020) 073006,
doi:10.1103/PhysRevD.102.073006,
[arXiv:2010.14402 [hep-ph]].

\bibitem{11} M.~Baouahi, \textit{et al.}, 
%``Laser-assisted kaon decay and CPT symmetry violation,'' 
Laser Phys. Lett. \textbf{18} (2021) 106001,
[arXiv:2107.11788 [hep-ph]].

\bibitem{12}
G.~Aad \textit{et al.}, [ATLAS],
%``Observation of a new particle in the search for the Standard Model Higgs boson with the ATLAS detector at the LHC,''
Phys. Lett. B \textbf{716} (2012) 1, 
doi:10.1016/j.physletb.2012.08.020,
[arXiv:1207.7214 [hep-ex]].

\bibitem{13}S.~Chatrchyan \textit{et al.} [CMS],
%``Observation of a New Boson at a Mass of 125 GeV with the CMS Experiment at the LHC,''
Phys. Lett. B \textbf{716} (2012) 30,
doi:10.1016/j.physletb.2012.08.021
[arXiv:1207.7235 [hep-ex]].

\bibitem{14}S.~Chatrchyan \textit{et al.} [CMS],
%``Measurement of the Properties of a Higgs Boson in the Four-Lepton Final State,''
Phys. Rev. D \textbf{89} (2014) 092007,
doi:10.1103/PhysRevD.89.092007,
[arXiv:1312.5353 [hep-ex]].

\bibitem{15}G.~Aad \textit{et al.} [ATLAS and CMS],
%``Combined Measurement of the Higgs Boson Mass in $pp$ Collisions at $\sqrt{s}=7$ and 8 TeV with the ATLAS and CMS Experiments,''
Phys. Rev. Lett. \textbf{114} (2015) 191803,
doi:10.1103/PhysRevLett.114.191803,
[arXiv:1503.07589 [hep-ex]].

\bibitem{16}V.~Khachatryan \textit{et al.} [CMS],
%``Search for new phenomena in monophoton final states in proton-proton collisions at $\sqrt s =$ 8 TeV,''
Phys. Lett. B \textbf{755} (2016) 102,
doi:10.1016/j.physletb.2016.01.057,
[arXiv:1410.8812 [hep-ex]].

\bibitem{17}I.~Tsukerman [ATLAS and CMS],
%``Measurements of the Higgs boson by ATLAS and CMS,''
J. Phys. Conf. Ser. \textbf{1390} (2019) 012030,
doi:10.1088/1742-6596/1390/1/012030.

\bibitem{18}M.~Aaboud \textit{et al.} [ATLAS],
%``Search for resonant $WZ$ production in the fully leptonic final state in proton-proton collisions at $\sqrt{s} = 13$ TeV with the ATLAS detector,''
Phys. Lett. B \textbf{787} (2018) 68,
doi:10.1016/j.physletb.2018.10.021,
[arXiv:1806.01532 [hep-ex]].

\bibitem{19}Takuya Morozumi, and Kotaro Tamai, Prog. Theor. Exp. Phys. \textbf{2014} (2014) 049201,
 https://doi.org/10.1093/ptep/ptu042.

\bibitem{20} V.V. Vien, Chin. J. Phys. \textbf{73} (2021) 47, https://doi.org/10.1016/j.cjph.2021.06.005.
\bibitem{21}M. Hashemi, Commun. Theor. Phys. \textbf{61} (2014) 69, https://doi.org/10.1088/0253-6102/61/1/11.
\bibitem{22}A. Djouadi, J.  Kalinowski, and P.M. Zerwas, Z. Phys. C \textbf{57} (1993) 569, https://doi.org/10.1007/BF01561476.

\bibitem{23}Jaume Guasch, Wolfgang Hollik, Arnd Kraft, Nuclear Physics B \textbf{596} (2001) 66, DOI: https://doi.org/10.1016/S0550-3213(00)00723-9.

\bibitem{24}Arbey, A., Mahmoudi, F., Stål, O, \textit{et al.}, Eur. Phys. J. C \textbf{78} (2018) 182, https://doi.org/10.1140/epjc/s10052-018-5651-1.

\bibitem{25}M. Aoki, S. Kanemura and H. Yokoya, Phys. Lett. B \textbf{725} (2013) 302, https://doi.org/10.1016/j.physletb.2013.07.011.

\bibitem{26}M. Gustafsson, S. Rydbeck, L. Lopez-Honorez and E. Lundstrom, Phys. Rev. D \textbf{86} (2012) 075019, 
https://doi.org/10.1103/PhysRevD.86.075019.

\bibitem{27}
M.~Krawczyk, D.~Sokolowska, P.~Swaczyna and B.~Swiezewska,
%``Constraining Inert Dark Matter by $R_{\gamma\gamma}$ and WMAP data,''
JHEP \textbf{09} (2013) 055,
doi:10.1007/JHEP09(2013)055,
[arXiv:1305.6266 [hep-ph]].

\bibitem{28}
M.~A.~D\'\i{}az, B.~Koch and S.~Urrutia-Quiroga,
%``Constraints to Dark Matter from Inert Higgs Doublet Model,''
Adv. High Energy Phys. \textbf{2016} (2016) 8278375,
doi:10.1155/2016/8278375,
[arXiv:1511.04429 [hep-ph]].

\bibitem{29}
K.~P.~Modak and D.~Majumdar,
%``Confronting Galactic and Extragalactic $\gamma$-rays Observed by Fermi-lat With Annihilating Dark Matter in an Inert Higgs Doublet Model,''
Astrophys. J. Suppl. \textbf{219} (2015) 37,
doi:10.1088/0067-0049/219/2/37,
[arXiv:1502.05682 [hep-ph]].

\bibitem{30}
L.~Linssen, A.~Miyamoto, M.~Stanitzki and H.~Weerts,
\textit{`Physics and Detectors at CLIC: CLIC Conceptual Design Report,''}
doi:10.5170/CERN-2012-003,
[arXiv:1202.5940 [physics.ins-det]].

\bibitem{31}
H.~Baer, T.~Barklow, K.~Fujii, Y.~Gao, A.~Hoang, S.~Kanemura, J.~List, H.~E.~Logan, A.~Nomerotski and M.~Perelstein, \textit{et al.}
\textit{``The International Linear Collider Technical Design Report - Volume 2: Physics,''}
[arXiv:1306.6352 [hep-ph]].

\bibitem{32}
J.~A.~Aguilar-Saavedra \textit{et al.} [ECFA/DESY LC Physics Working Group],
\textit{``TESLA: The Superconducting electron positron linear collider with an integrated x-ray laser laboratory. Technical design report. Part 3. Physics at an e+ e- linear collider,''}
[arXiv:hep-ph/0106315 [hep-ph]].

\bibitem{33}
J.~B.~Guimar\~aes da Costa \textit{et al.} [CEPC Study Group],
\textit{``CEPC Conceptual Design Report: Volume 2 - Physics \& Detector,''},
[arXiv:1811.10545 [hep-ex]].

\bibitem{34}
 [CEPC Study Group],
\textit{``CEPC Conceptual Design Report: Volume 1 - Accelerator,''},
[arXiv:1809.00285 [physics.acc-ph]].

\bibitem{35}D.~M.~Volkov,
%``Uber eine Klasse von Losungen der Diracschen Gleichung,''
Z. Phys. \textbf{94} (1935) 250,
doi:10.1007/BF01331022.

\bibitem{36}W. Greiner and B. Mueller, Gauge Theory of Weak Interactions, 3rd ed. (Springer, Berlin, 2000).

\bibitem{37}Yao-Bei Liu, Hong-Mei Han, Xue-Lei Wang, Eur. Phys. J. C \textbf{53} (2008) 615, https://doi.org/10.1140/epjc/s10052-007-0492-3.

\bibitem{38} P. A. Zyla, \textit{et al}, (Particle Data Group), Prog. Theor. Exp. Phys. \textbf{2020} (2020) 083C01, https://doi.org/10.1093/ptep/ptaa104.

\bibitem{39}S. Heinemeyer, C. Schappacher,  Eur. Phys. J. C \textbf{76} (2016) 535, https://doi.org/10.1140/epjc/s10052-016-4383-3.

\bibitem{40}F. V. Bunkin and M. V. Fedorov, Sov. Phys. JETP \textbf{22} (1966) 844 ; N. M. Kroll and K. M. Watson, Phys. Rev. A \textbf{8} (1973) 804.
\end{thebibliography}
\end{document}